\documentclass[12pt]{article}
\usepackage{epsfig}
\usepackage{amsmath}
\usepackage{hhline}
\usepackage{amssymb}
\usepackage{times}
\usepackage{cite}
\usepackage{colordvi}
\usepackage{captcont}

\newlength{\dinwidth}
\newlength{\dinmargin}
\begin{document}  
\newcommand{\pom}{{I\!\!P}}
\newcommand{\reg}{{I\!\!R}}
\newcommand{\slowpi}{\pi_{\mathit{slow}}}
\newcommand{\fiidiii}{F_2^{D(3)}}
\newcommand{\fiidiiiarg}{\fiidiii\,(\beta,\,Q^2,\,x)}
\newcommand{\n}{1.19\pm 0.06 (stat.) \pm0.07 (syst.)}
\newcommand{\nz}{1.30\pm 0.08 (stat.)^{+0.08}_{-0.14} (syst.)}
\newcommand{\fiidiiiful}{F_2^{D(4)}\,(\beta,\,Q^2,\,x,\,t)}
\newcommand{\fiipom}{\tilde F_2^D}
\newcommand{\ALPHA}{1.10\pm0.03 (stat.) \pm0.04 (syst.)}
\newcommand{\ALPHAZ}{1.15\pm0.04 (stat.)^{+0.04}_{-0.07} (syst.)}
\newcommand{\fiipomarg}{\fiipom\,(\beta,\,Q^2)}
\newcommand{\pomflux}{f_{\pom / p}}
\newcommand{\nxpom}{1.19\pm 0.06 (stat.) \pm0.07 (syst.)}
\newcommand {\gapprox}
   {\raisebox{-0.7ex}{$\stackrel {\textstyle>}{\sim}$}}
\newcommand {\lapprox}
   {\raisebox{-0.7ex}{$\stackrel {\textstyle<}{\sim}$}}
\def\gsim{\,\lower.25ex\hbox{$\scriptstyle\sim$}\kern-1.30ex%
\raise 0.55ex\hbox{$\scriptstyle >$}\,}
\def\lsim{\,\lower.25ex\hbox{$\scriptstyle\sim$}\kern-1.30ex%
\raise 0.55ex\hbox{$\scriptstyle <$}\,}
\newcommand{\pomfluxarg}{f_{\pom / p}\,(x_\pom)}
\newcommand{\dsf}{\mbox{$F_2^{D(3)}$}}
\newcommand{\dsfva}{\mbox{$F_2^{D(3)}(\beta,Q^2,x_{I\!\!P})$}}
\newcommand{\dsfvb}{\mbox{$F_2^{D(3)}(\beta,Q^2,x)$}}
\newcommand{\dsfpom}{$F_2^{I\!\!P}$}
\newcommand{\gap}{\stackrel{>}{\sim}}
\newcommand{\lap}{\stackrel{<}{\sim}}
\newcommand{\fem}{$F_2^{em}$}
\newcommand{\tsnmp}{$\tilde{\sigma}_{NC}(e^{\mp})$}
\newcommand{\tsnm}{$\tilde{\sigma}_{NC}(e^-)$}
\newcommand{\tsnp}{$\tilde{\sigma}_{NC}(e^+)$}
\newcommand{\st}{$\star$}
\newcommand{\sst}{$\star \star$}
\newcommand{\ssst}{$\star \star \star$}
\newcommand{\sssst}{$\star \star \star \star$}
\newcommand{\tw}{\theta_W}
\newcommand{\sw}{\sin{\theta_W}}
\newcommand{\cw}{\cos{\theta_W}}
\newcommand{\sww}{\sin^2{\theta_W}}
\newcommand{\cww}{\cos^2{\theta_W}}
\newcommand{\trm}{m_{\perp}}
\newcommand{\trp}{p_{\perp}}
\newcommand{\trmm}{m_{\perp}^2}
\newcommand{\trpp}{p_{\perp}^2}
\newcommand{\alp}{\alpha_s}

\newcommand{\alps}{\alpha_s}
\newcommand{\sqrts}{$\sqrt{s}$}
\newcommand{\LO}{$O(\alpha_s^0)$}
\newcommand{\Oa}{$O(\alpha_s)$}
\newcommand{\Oaa}{$O(\alpha_s^2)$}
\newcommand{\PT}{p_{\perp}}
\newcommand{\JPSI}{J/\psi}
\newcommand{\sh}{\hat{s}}
\newcommand{\uh}{\hat{u}}
\newcommand{\MP}{m_{J/\psi}}
\newcommand{\PO}{I\!\!P}
\newcommand{\xbj}{x}
\newcommand{\xpom}{x_{\PO}}
\newcommand{\ttbs}{\char'134}
\newcommand{\xpomlo}{3\times10^{-4}}  
\newcommand{\xpomup}{0.05}  
\newcommand{\dgr}{^\circ}
\newcommand{\pbarnt}{\,\mbox{{\rm pb$^{-1}$}}}
\newcommand{\gev}{\,\mbox{GeV}}
\newcommand{\WBoson}{\mbox{$W$}}
\newcommand{\fbarn}{\,\mbox{{\rm fb}}}
\newcommand{\fbarnt}{\,\mbox{{\rm fb$^{-1}$}}}
\newcommand{\dsdx}[1]{$d\sigma\!/\!d #1\,$}
\newcommand{\eV}{\mbox{e\hspace{-0.08em}V}}
%
%
\newcommand{\qsq}{\ensuremath{Q^2} }
\newcommand{\gevsq}{\ensuremath{\mathrm{GeV}^2} }
\newcommand{\et}{\ensuremath{E_t^*} }
\newcommand{\rap}{\ensuremath{\eta^*} }
\newcommand{\gp}{\ensuremath{\gamma^*}p }
\newcommand{\dsiget}{\ensuremath{{\rm d}\sigma_{ep}/{\rm d}E_t^*} }
\newcommand{\dsigrap}{\ensuremath{{\rm d}\sigma_{ep}/{\rm d}\eta^*} }

\newcommand{\dstar}{\ensuremath{D^*}}
\newcommand{\dstarp}{\ensuremath{D^{*+}}}
\newcommand{\dstarm}{\ensuremath{D^{*-}}}
\newcommand{\dstarpm}{\ensuremath{D^{*\pm}}}
\newcommand{\zDs}{\ensuremath{z(\dstar )}}
\newcommand{\Wgp}{\ensuremath{W_{\gamma p}}}
\newcommand{\ptds}{\ensuremath{p_t(\dstar )}}
\newcommand{\etads}{\ensuremath{\eta(\dstar )}}
\newcommand{\ptj}{\ensuremath{p_t(\mbox{jet})}}
\newcommand{\ptjn}[1]{\ensuremath{p_t(\mbox{jet$_{#1}$})}}
\newcommand{\etaj}{\ensuremath{\eta(\mbox{jet})}}
\newcommand{\detadsj}{\ensuremath{\eta(\dstar )\, \mbox{-}\, \etaj}}

\def\Journal#1#2#3#4{{#1} {\bf #2} (#3) #4}
\def\NCA{\em Nuovo Cimento}
\def\NIM{\em Nucl. Instrum. Methods}
\def\NIMA{{\em Nucl. Instrum. Methods} {\bf A}}
\def\NPB{{\em Nucl. Phys.}   {\bf B}}
\def\PLB{{\em Phys. Lett.}   {\bf B}}
\def\PRL{\em Phys. Rev. Lett.}
\def\PRD{{\em Phys. Rev.}    {\bf D}}
\def\ZPC{{\em Z. Phys.}      {\bf C}}
\def\EJC{{\em Eur. Phys. J.} {\bf C}}
\def\CPC{\em Comp. Phys. Commun.}

\newcommand{\be}{\begin{equation}} 
\newcommand{\ee}{\end{equation}} 
\newcommand{\ba}{\begin{eqnarray}} 
\newcommand{\ea}{\end{eqnarray}} 

\def\ra{\rightarrow}
\def\Q2{\mbox{$Q^2$}}
\def\xbj{\mbox{$x_{Bj}$}}
\def\Pom{\mbox{I$\!$P}}
\def\Pem{\mbox{$I\!\!P$}}
\def\Pma{I\!\!P}
\def\xip{\mbox{$x_{\Pma}$}}
\def\pbi{pb$^-1$}
\def\chisq{\chi^2}

\begin{titlepage}

\noindent
\begin{flushleft}
{\tt DESY 12-100 \hfill ISSN 0418-9833} \\
{\tt June 2012}                  \\
\end{flushleft}

\noindent

\vspace{2cm}
\begin{center}
\begin{Large}

{\bf Combined inclusive diffractive cross sections measured with 
forward proton spectrometers in deep inelastic {\boldmath $ep$} scattering 
at HERA\\}

\vspace{2cm}

H1 and ZEUS Collaborations

\end{Large}
\end{center}

\vspace{2cm}

\begin{abstract}
  A combination of the inclusive diffractive cross section measurements made by the H1 and ZEUS Collaborations at HERA is 
 presented. 
 The analysis uses samples of diffractive deep inelastic $ep$ scattering data at a centre-of-mass energy $\sqrt{s}$~=~318 
GeV where leading protons  are 
  detected by dedicated spectrometers. Correlations of systematic uncertainties are taken into account, 
 resulting in an improved precision of the cross section measurement which reaches 6\% for the most precise points.
 The combined data cover the range $2.5 < Q^2 < 200$ GeV$^2$ in photon virtuality, $0.00035 < 
 \xip < 0.09$ in proton fractional 
 momentum loss, $0.09 < |t| < 0.55$ GeV$^2$ in squared four-momentum transfer at the proton vertex and $0.0018 < \beta < 0.816$ 
 in $\beta=x/\xip$, where $x$ is the Bjorken scaling variable.

\end{abstract}

\vspace{1.5cm}

\begin{center}
  Submitted to \EJC
\end{center}

\end{titlepage}

\begin{flushleft}
  F.D.~Aaron$^{13,a4}$,           
H.~Abramowicz$^{71, a45}$, 
I.~Abt$^{56}$, 
L.~Adamczyk$^{35}$, 
M.~Adamus$^{84}$, 
R.~Aggarwal$^{14, a12}$, 
C.~Alexa$^{13}$,                
V.~Andreev$^{53}$,             
S.~Antonelli$^{10}$, 
P.~Antonioli$^{9}$
% ...

\bigskip{\it

   $^{1}$ I. Physikalisches Institut der RWTH, Aachen, Germany \\
 $^{2}$ {\it Institute of Physics and Technology of Ministry of Education and Science of Kazakhstan, Almaty, Kazakhstan}\\
 $^{3}$ {\it NIKHEF and University of Amsterdam, Amsterdam, Netherlands}~$^{b29}$\\
   $^{4}$ Inter-University Institute for High Energies ULB-VUB, Brussels and         Universiteit Antwerpen, Antwerpen, Belgium~$^{b2}$ \\
 $^{5}$ {\it Argonne National Laboratory, Argonne, Illinois 60439-4815, USA}~$^{b13}$\\
   $^{6}$ Vinca Institute of Nuclear Sciences, University of Belgrade,          1100 Belgrade, Serbia \\
 $^{7}$ {\it Andrews University, Berrien Springs, Michigan 49104-0380, USA}\\
   $^{8}$ School of Physics and Astronomy, University of Birmingham,          Birmingham, UK~$^{b16}$ \\
 $^{9}$ {\it INFN Bologna, Bologna, Italy}~$^{b14}$\\
 $^{10}$ {\it University and INFN Bologna, Bologna, Italy}~$^{b14}$\\
 $^{11}$ {\it Physikalisches Institut der Universit\"at Bonn, Bonn, Germany}~$^{b15}$\\
 $^{12}$ {\it H.H.~Wills Physics Laboratory, University of Bristol, Bristol, United Kingdom}~$^{b16}$\\
   $^{13}$ National Institute for Physics and Nuclear Engineering (NIPNE) ,          Bucharest, Romania~$^{b11}$ \\
 $^{14}$ {\it Panjab University, Department of Physics, Chandigarh, India}\\
 $^{15}$ {\it Department of Engineering in Management and Finance, Univ. of the Aegean, Chios, Greece}\\
 $^{16}$ {\it Physics Department, Ohio State University, Columbus, Ohio 43210, USA}~$^{b13}$\\
 $^{17}$ {\it Calabria University, Physics Department and INFN, Cosenza, Italy}~$^{b14}$\\
   $^{18}$ The Henryk Niewodniczanski Institute of Nuclear Physics, Polish Academy of Sciences, Cracow, Poland~$^{b4}$ \\
 $^{19}$ {\it Department of Physics, Jagellonian University, Cracow, Poland}\\
 $^{20}$ {\it Kyungpook National University, Center for High Energy Physics, Daegu, South Korea}~$^{b23}$\\
   $^{21}$ Institut f\"ur Physik, TU Dortmund, Dortmund, Germany~$^{b1}$ \\
   $^{22}$ Joint Institute for Nuclear Research, Dubna, Russia \\
 $^{23}$ {\it INFN Florence, Florence, Italy}~$^{b14}$\\
 $^{24}$ {\it University and INFN Florence, Florence, Italy}~$^{b14}$\\
 $^{25}$ {\it Fakult\"at f\"ur Physik der Universit\"at Freiburg i.Br., Freiburg i.Br., Germany}\\
   $^{26}$ CEA, DSM/Irfu, CE-Saclay, Gif-sur-Yvette, France \\
 $^{27}$ {\it School of Physics and Astronomy, University of Glasgow, Glasgow, United Kingdom}~$^{b16}$\\
$^{28}$ Institut f\"ur Experimentalphysik, Universit\"at Hamburg, Hamburg, Germany~$^{b1}$$^{,}$$^{b21}$\\ 
 $^{29}$ {\it Deutsches Elektronen-Synchrotron DESY, Hamburg, Germany}\\
   $^{30}$ Physikalisches Institut, Universit\"at Heidelberg,          Heidelberg, Germany~$^{b1}$ \\
   $^{31}$ Kirchhoff-Institut f\"ur Physik, Universit\"at Heidelberg,          Heidelberg, Germany~$^{b1}$ \\
   $^{32}$ Max-Planck-Institut f\"ur Kernphysik, Heidelberg, Germany \\
 $^{33}$ {\it Nevis Laboratories, Columbia University, Irvington on Hudson, New York 10027, USA}~$^{b18}$\\
   $^{34}$ Institute of Experimental Physics, Slovak Academy of          Sciences, Ko\v{s}ice, Slovak Republic~$^{b5}$ \\
 $^{35}$ {\it AGH-University of Science and Technology, Faculty of Physics and Applied Computer Science, Krakow, Poland}~$^{b20}$\\
 $^{36}$ {\it Jabatan Fizik, Universiti Malaya, 50603 Kuala Lumpur, Malaysia}~$^{b17}$\\
 $^{37}$ {\it Institute for Universe and Elementary Particles, Chonnam National University, Kwangju, South Korea}\\
 $^{38}$ {\it Institute for Nuclear Research, National Academy of Sciences, Kyiv, Ukraine}\\
 $^{39}$ {\it Department of Nuclear Physics, National Taras Shevchenko University of Kyiv, Kyiv, Ukraine}\\
   $^{40}$ Department of Physics, University of Lancaster,          Lancaster, UK~$^{b16}$ \\
   $^{41}$ Department of Physics, University of Liverpool,         Liverpool, UK~$^{b16}$ \\
   $^{42}$ School of Physics and Astronomy, Queen Mary, University of London,          London, UK~$^{b16}$ \\
 $^{43}$ {\it Imperial College London, High Energy Nuclear Physics Group, London, United Kingdom}~$^{b16}$\\
 $^{44}$ {\it Physics and Astronomy Department, University College London, London, United Kingdom}~$^{b16}$\\
 $^{45}$ {\it Institut de Physique Nucl\'{a14}aire, Universit\'{a14} Catholique de Louvain, Louvain-la-Neuve,\\ Belgium}~$^{b24}$\\
   $^{46}$ Physics Department, University of Lund,          Lund, Sweden~$^{b6}$ \\
   $^{47}$ Departamento de Fisica, CINVESTAV  IPN, M\'exico City, M\'exico~$^{b9}$ \\
 $^{48}$ {\it Department of Physics, University of Wisconsin, Madison, Wisconsin 53706, USA}~$^{b13}$\\
 $^{49}$ {\it Departamento de F\'{\i}sica Te\'orica, Universidad Aut\'onoma de Madrid, Madrid, Spain}~$^{b25}$\\
   $^{50}$ CPPM, Aix-Marseille Univ, CNRS/IN2P3, 13288 Marseille, France \\
 $^{51}$ {\it Department of Physics, McGill University, Montr\'eal, Qu\'ebec, Canada H3A 2T8}~$^{b26}$\\
   $^{52}$ Institute for Theoretical and Experimental Physics, Moscow, Russia~$^{b10}$ \\
   $^{53}$ Lebedev Physical Institute, Moscow, Russia \\
 $^{54}$ {\it Moscow Engineering Physics Institute, Moscow, Russia}~$^{b27}$\\
 $^{55}$ {\it Lomonosov Moscow State University, Skobeltsyn Institute of Nuclear Physics, Moscow, Russia}~$^{b28}$\\
 $^{56}$ {\it Max-Planck-Institut f\"ur Physik, M\"unchen, Germany}\\
   $^{57}$ LAL, Universit\'e Paris-Sud, CNRS/IN2P3, Orsay, France \\
 $^{58}$ {\it Department of Physics, University of Oxford, Oxford, United Kingdom}~$^{b16}$\\
   $^{59}$ STFC, Rutherford Appleton Laboratory, Didcot, Oxfordshire, UK~$^{b16}$ \\
 $^{60}$ {\it INFN Padova, Padova, Italy}~$^{b14}$\\
 $^{61}$ {\it Dipartimento di Fisica dell' Universit\`a and INFN, Padova, Italy}~$^{b14}$\\
   $^{62}$ LLR, Ecole Polytechnique, CNRS/IN2P3, Palaiseau, France \\
   $^{63}$ LPNHE, Universit\'e Pierre et Marie Curie Paris 6,   Universit\'e Denis Diderot Paris 7, CNRS/IN2P3, Paris, France \\
   $^{64}$ Faculty of Science University of Montenegro, Podgorica, Montenegro~$^{b12}$ \\
   $^{65}$ Institute of Physics of the Academy of Sciences of the Czech Republic,  Praha, Czech Republic~$^{b7}$ \\
   $^{66}$ Faculty of Mathematics and Physics of Charles University, Praha, Czech Republic~$^{b7}$ \\
 $^{67}$ {\it Department of Particle Physics and Astrophysics, Weizmann Institute, Rehovot, Israel}\\
   $^{68}$ Dipartimento di Fisica Universit\`a di Roma Tre   and INFN Roma~3, Roma, Italy \\
 $^{69}$ {\it Dipartimento di Fisica, Universit\`a 'La Sapienza' and INFN, Rome, Italy}~$^{b14}$\\
   $^{70}$ Institute for Nuclear Research and Nuclear Energy, Sofia, Bulgaria \\
 $^{71}$ {\it Raymond and Beverly Sackler Faculty of Exact Sciences, School of Physics, Tel Aviv University, Tel Aviv, Israel}~$^{b30}$\\
 $^{72}$ {\it Polytechnic University, Tokyo, Japan}~$^{b22}$\\
 $^{73}$ {\it Department of Physics, Tokyo Institute of Technology, Tokyo, Japan}~$^{b22}$\\
 $^{74}$ {\it Department of Physics, University of Tokyo, Tokyo, Japan}~$^{b22}$\\
 $^{75}$ {\it Tokyo Metropolitan University, Department of Physics, Tokyo, Japan}~$^{b22}$\\
 $^{76}$ {\it Universit\`a di Torino and INFN, Torino, Italy}~$^{b14}$\\
 $^{77}$ {\it Universit\`a del Piemonte Orientale, Novara, and INFN, Torino, Italy}~$^{b14}$\\
 $^{78}$ {\it Department of Physics, University of Toronto, Toronto, Ontario, Canada M5S 1A7}~$^{b26}$\\
 $^{79}$ {\it Institute of Particle and Nuclear Studies, KEK, Tsukuba, Japan}~$^{b22}$\\
   $^{80}$ Institute of Physics and Technology of the Mongolian  Academy of Sciences, Ulaanbaatar, Mongolia \\
 $^{81}$ {\it Department of Physics, Pennsylvania State University, University Park, Pennsylvania 16802, USA}~$^{b18}$\\
   $^{82}$ Paul Scherrer Institut,    Villigen, Switzerland \\
 $^{83}$ {\it Faculty of Physics, University of Warsaw, Warsaw, Poland}\\
 $^{84}$ {\it National Centre for Nuclear Research, Warsaw, Poland}\\
   $^{85}$ Fachbereich C, Universit\"at Wuppertal, Wuppertal, Germany \\
   $^{86}$ Yerevan Physics Institute, Yerevan, Armenia \\
 $^{87}$ {\it Meiji Gakuin University, Faculty of General Education, Yokohama, Japan}~$^{b22}$\\
 $^{88}$ {\it Department of Physics, York University, Ontario, Canada M3J 1P3}~$^{b26}$\\
   $^{89}$ Departamento de Fisica Aplicada,        CINVESTAV, M\'erida, Yucat\'an, M\'exico~$^{b9}$ \\
 $^{90}$ {\it Deutsches Elektronen-Synchrotron DESY, Zeuthen, Germany}\\
   $^{91}$ Institut f\"ur Teilchenphysik, ETH, Z\"urich, Switzerland~$^{b8}$ \\
   $^{92}$ Physik-Institut der Universit\"at Z\"urich, Z\"urich, Switzerland~$^{b8}$ \\

\bigskip
 $ ^{a1}$ Also at Rechenzentrum, Universit\"at Wuppertal,
          Wuppertal, Germany \\
 $ ^{a2}$ Also at IPNL, Universit\'e Claude Bernard Lyon 1, CNRS/IN2P3,
          Villeurbanne, France \\
 $ ^{a3}$ Also at CERN, Geneva, Switzerland \\
 $ ^{a4}$ Also at Faculty of Physics, University of Bucharest,
          Bucharest, Romania \\
 $ ^{a5}$ Also at Ulaanbaatar University, Ulaanbaatar, Mongolia \\
 $ ^{a6}$ Supported by the Initiative and Networking Fund of the
          Helmholtz Association (HGF) under the contract VH-NG-401 and S0-072. \\
 $ ^{a7}$ Absent on leave from NIPNE-HH, Bucharest, Romania \\
 $ ^{a8}$ Also at  Department of Physics, University of Toronto,
          Toronto, Ontario, Canada M5S 1A7 \\
 $ ^{a9}$ Also at LAPP, Universit\'e de Savoie, CNRS/IN2P3,
          Annecy-le-Vieux, France \\
$^{a10}$ Now at University of Salerno, Italy\\
$^{a11}$ Now at Queen Mary University of London, United Kingdom\\
$^{a12}$ Also funded by Max Planck Institute for Physics, Munich, Germany\\
$^{a13}$ Also Senior Alexander von Humboldt Research Fellow at Hamburg University, Institute of Experimental Physics, Hamburg, Germany\\
$^{a14}$ Also at Cracow University of Technology, Faculty of Physics, Mathemathics and Applied Computer Science, Poland\\
$^{a15}$ Supported by the research grant No. 1 P03B 04529 (2005-2008)\\
$^{a16}$ Supported by the Polish National Science Centre, project No. DEC-2011/01/BST2/03643\\
$^{a17}$ Now at Rockefeller University, New York, NY 10065, USA\\
$^{a18}$ Now at DESY group FS-CFEL-1\\
$^{a19}$ Now at Institute of High Energy Physics, Beijing, China\\
$^{a20}$ Now at DESY group FEB, Hamburg, Germany\\
$^{a21}$ Also at Moscow State University, Russia\\
$^{a22}$ Now at University of Liverpool, United Kingdom\\
$^{a23}$ Now at CERN, Geneva, Switzerland\\
$^{a24}$ Also affiliated with Universtiy College London, UK\\
$^{a25}$ Now at Goldman Sachs, London, UK\\
$^{a26}$ Also at Institute of Theoretical and Experimental Physics, Moscow, Russia\\
$^{a27}$ Also at FPACS, AGH-UST, Cracow, Poland\\
$^{a28}$ Partially supported by Warsaw University, Poland\\
$^{a29}$ Now at Istituto Nucleare di Fisica Nazionale (INFN), Pisa, Italy\\
$^{a30}$ Now at Haase Energie Technik AG, Neum\"unster, Germany\\
$^{a31}$ Now at Department of Physics, University of Bonn, Germany\\
$^{a32}$ Now at Biodiversit\"at und Klimaforschungszentrum (BiK-F), Frankfurt, Germany\\
$^{a33}$ Also affiliated with DESY, Germany\\
$^{a34}$ Also at University of Tokyo, Japan\\
$^{a35}$ Now at Kobe University, Japan\\
$^{a37}$ Supported by DESY, Germany\\
$^{a38}$ Member of National Technical University of Ukraine, Kyiv Polytechnic Institute,  Kyiv, Ukraine\\
$^{a39}$ Member of National University of Kyiv - Mohyla Academy, Kyiv, Ukraine\\
$^{a40}$ Partly supported by the Russian Foundation for Basic Research, grant 11-02-91345-DFG\_a\\
$^{a41}$ Alexander von Humboldt Professor; also at DESY and University of Oxford\\
$^{a42}$ STFC Advanced Fellow\\
$^{a43}$ Now at LNF, Frascati, Italy\\
$^{a44}$ This material was based on work supported by the National Science Foundation, while working at the Foundation.\\
$^{a45}$ Also at Max Planck Institute for Physics, Munich, Germany, External Scientific Member\\
$^{a46}$ Now at Tokyo Metropolitan University, Japan\\
$^{a47}$ Now at Nihon Institute of Medical Science, Japan\\
$^{a48}$ Now at Osaka University, Osaka, Japan\\
$^{a49}$ Also at Lodz University, Poland\\ 
$^{a50}$ Member of Lodz  University, Poland\\ 
$^{a51}$ Now at Department of Physics, Stockholm University, Stockholm, Sweden\\
$^{a52}$ Also at Cardinal Stefan Wyszy\'nski University, Warsaw, Poland\\

\bigskip

 $^{b1}$ Supported by the Bundesministerium f\"ur Bildung und Forschung, FRG, under contract numbers 05H09GUF, 05H09VHC, 05H09VHF,  05H16PEA \\
 $^{b2}$ Supported by FNRS-FWO-Vlaanderen, IISN-IIKW and IWT  and  by Interuniversity Attraction Poles Programme,  Belgian Science Policy \\
 $^{b4}$ Supported by Polish Ministry of Science and Higher  Education, grants  DPN/N168/DESY/2009 and DPN/N188/DESY/2009 \\
 $^{b5}$ Supported by VEGA SR grant no. 2/7062/ 27 \\
 $^{b6}$ Supported by the Swedish Natural Science Research Council \\
 $^{b7}$ Supported by the Ministry of Education of the Czech Republic under the projects  LC527, INGO-LA09042 and   MSM0021620859 \\
 $^{b8}$ Supported by the Swiss National Science Foundation \\
 $^{b9}$ Supported by  CONACYT,  M\'exico, grant 48778-F \\
 $^{b10}$ Russian Foundation for Basic Research (RFBR), grant no 1329.2008.2 and Rosatom \\
 $^{b11}$ Supported by the Romanian National Authority for Scientific Research  under the contract PN 09370101 \\
 $^{b12}$ Partially Supported by Ministry of Science of Montenegro,  no. 05-1/3-3352 \\
$^{b13}$  Supported by the US Department of Energy\\
$^{b14}$  Supported by the Italian National Institute for Nuclear Physics (INFN) \\
$^{b15}$  Supported by the German Federal Ministry for Education and Research (BMBF), under contract No. 05 H09PDF\\
$^{b16}$  Supported by the Science and Technology Facilities Council, UK\\
$^{b17}$  Supported by an FRGS grant from the Malaysian government\\
$^{b18}$  Supported by the US National Science Foundation. Any opinion, findings and conclusions or recommendations expressed in this material are those of the authors and do not necessarily reflect the views of the  National Science Foundation.\\
$^{b20}$  Supported by the Polish Ministry of Science and Higher Education and its grants for Scientific Research\\
$^{b21}$  Supported by the German Federal Ministry for Education and Research (BMBF), under contract No. 05h09GUF, and the SFB 676 of the Deutsche Forschungsgemeinschaft (DFG) \\
$^{b22}$  Supported by the Japanese Ministry of Education, Culture, Sports, Science and Technology (MEXT) and its grants for Scientific Research\\
$^{b23}$  Supported by the Korean Ministry of Education and Korea Science and Engineering Foundation\\
$^{b24}$  Supported by FNRS and its associated funds (IISN and FRIA) and by an Inter-University Attraction Poles Programme subsidised by the Belgian Federal Science Policy Office\\
$^{b25}$  Supported by the Spanish Ministry of Education and Science through funds provided by CICYT\\
$^{b26}$  Supported by the Natural Sciences and Engineering Research Council of Canada (NSERC) \\
$^{b27}$  Partially supported by the German Federal Ministry for Education and Research (BMBF)\\
$^{b28}$  Supported by RF Presidential grant N 4142.2010.2 for Leading Scientific Schools, by the Russian Ministry of Education and Science through its grant for Scientific Research on High Energy Physics and under contract No.02.740.11.0244 \\
$^{b29}$  Supported by the Netherlands Foundation for Research on Matter (FOM)\\
$^{b30}$  Supported by the Israel Science Foundation\\

\vspace{0.2cm}
$^{\dagger}$    deceased \\

}

\end{flushleft}

\newpage

\section{Introduction}
\label{sec-int}

Diffractive collisions in deep inelastic electron-proton 
scattering (DIS), $ep \to eXp$, have been studied extensively at the HERA collider.  
They can be viewed as resulting from processes in which the exchanged 
photon probes a colour-singlet combination of partons. The photon 
virtuality, $Q^2$, supplies a hard scale, which 
allows the application of perturbative quantum chromodynamics (QCD).
Diffractive reactions in DIS are a tool 
to investigate low-momentum partons in the proton, notably through the
study of diffractive parton distribution functions (DPDFs), determined by a QCD analysis of the data.

\begin{figure}[!h]
\vfill
\begin{center}
\hskip 3.0cm \includegraphics[scale = 0.33]{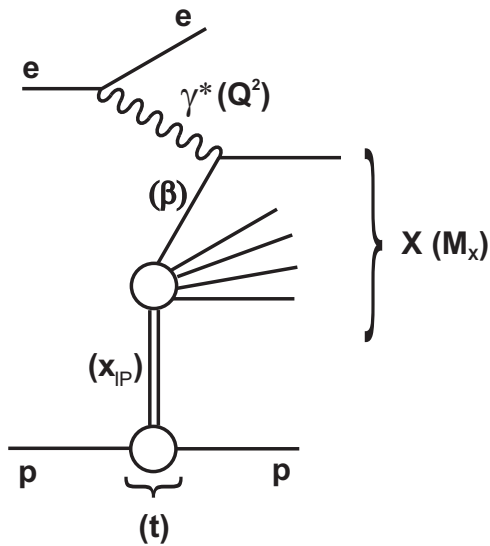}
\end{center}
\vspace{-0.3cm}

\caption{
Diagram of the reaction $ep \rightarrow eXp$.
}
\label{fig-diagram}
\vfill
\end{figure}

In diffractive $ep$ scattering the virtual photon dissociates 
at a photon-proton centre-of-mass energy $W$ and squared four-momentum transfer $t$ at 
the proton vertex (figure~\ref{fig-diagram}), producing a hadronic system 
$X$ with mass $M_X$.  
The fractional longitudinal
momentum loss of the proton is
denoted as $\xip$, while the fraction of this momentum 
taking part in the interaction with the photon is denoted as $\beta$. 
These variables are related to Bjorken $x$ by $x=\beta \, \xip$. 
The variable $\beta$ is related to $M_X$ and $Q^2$ by $\beta\simeq Q^2/(Q^2+M_X^2)$.
The variables $W, Q^2$ and the inelasticity $y$ are related by $W^2\simeq sy-Q^2$, where $s$ is the square of the $ep$ centre-of-mass energy.

Similarly to inclusive DIS,
diffractive cross section measurements 
are conventionally expressed in terms of the
reduced diffractive cross section, $\sigma_r^{D(4)}$, which is related
to the measured $ep$ cross section by
\begin{eqnarray}
\frac{{\rm d} \sigma^{ep \rightarrow eXp}}{{\rm d} \beta {\rm d} Q^2 {\rm d} \xip {\rm d} t} =
\frac{4\pi\alpha^2}{\beta Q^4} \ \ \left[1-y+\frac{y^2}{2}\right] \ \
\sigma_r^{D(4)}(\beta,Q^2,\xip,t) \ .
\label{sigma-2}
\end{eqnarray}
The reduced cross section $\sigma_r^{D(3)}(\beta,Q^2,\xip)$ is obtained by integrating $\sigma_r^{D(4)}(\beta,Q^2,\xip,t)$ over $t$.
At small and moderate values of the inelasticity $y$, $\sigma_r^{D(3)}$
is approximately equal
to the diffractive structure function $F_2^{D(3)}$ to good approximation.

Experimentally, diffractive $ep$ scattering is 
characterised by the presence of a leading proton in the final state 
carrying most of the proton beam energy and by a depletion of 
hadronic activity in the pseudo-rapidity\footnote{The pseudo-rapidity is defined as $\eta=-\ln\tan\theta/2$ where the 
polar angle $\theta$ is measured with respect to the proton beam direction.} distribution of particles 
(large rapidity gap, LRG) in the forward (proton) direction. 
Both of these signatures have been exploited in various analyses 
by H1 and ZEUS to select diffractive samples either  
by tagging the outgoing proton in dedicated proton 
spectrometers~\cite{h1-fps1,h1-fps2,zeus-lps1,zeus-lrglps}  
or by requiring the presence of a large rapidity 
gap~\cite{zeus-lrglps,h1-lrg1,h1-lrg2}. 
The two methods differ partially in the accessible kinematic ranges 
(higher $\xpom$ for the proton-tagged data) and substantially in their 
dominant sources of systematic uncertainties. 
In LRG-based measurements, 
 the largest uncertainty arises from proton dissociative events, 
 $ep \rightarrow eXN$, in which 
 the proton dissociates into a low mass state $N$. 
%
%
Low $\xip$ samples selected by the proton spectrometers 
have little or no proton dissociation contribution, 
but their precision is limited statistically by the  
small acceptances and systematically by large uncertainties in the proton
tagging efficiency, which strongly depends on the proton-beam optics.
The results from both methods are found to be 
consistent~\cite{h1-fps1,h1-fps2,zeus-lrglps,h1-lrg2,martapaul}.

 Combining  measurements can provide more precise and 
kinematically extended data than the individual measurements. 
In this paper, a combination of 
the H1~\cite{h1-fps2, h1-fps1} and the ZEUS~\cite{zeus-lrglps,zeus-lps1} proton spectrometer results is presented.
The combination is performed using the weighted averaging method 
introduced in~\cite{glazov} and extended in~\cite{:2009wt,Aaron:2009bp}. 
The correlated systematic uncertainties and global normalisations are constrained in the fit such that one coherent data set is 
obtained. Since H1 
and ZEUS have employed different experimental techniques, using different detectors and methods of kinematic reconstruction, the combination leads to significantly reduced uncertainties. 
 The kinematic range of the combined data is: $2.5 \leq Q^2 \leq 200$~GeV$^2$, $0.0018 \leq \beta \leq 0.816$, $0.00035 \leq 
 x_{\pom} \leq 0.09$ and $0.09 < |t| < 0.55$~GeV$^2$. The latter requirement restricts the analysis to the $t$ range directly accessible 
by both the H1 and ZEUS proton spectrometers. 


\section{Combination of the H1 and ZEUS measurements}

\subsection{Data samples}

The H1~\cite{H1_det} and ZEUS~\cite{ZEUS_det} detectors were general purpose instruments which consisted of 
tracking systems surrounded by electromagnetic and hadronic calorimeters and muon detectors, ensuring close to 4$\pi$ 
coverage about the $ep$ interaction point.
Both detectors were equipped with proton spectrometers; the Leading Proton 
Spectrometer (LPS) for ZEUS, the Forward Proton Spectrometer (FPS) and the 
Very Forward Proton Spectrometer (VFPS) for H1. These spectrometers were located $60$ to $220$~m away from the main 
detectors in the forward (proton beam) direction. 

The combination is based on the cross sections measured with the H1 FPS~\cite{h1-fps1,h1-fps2} and the ZEUS 
LPS~\cite{zeus-lps1,zeus-lrglps}. 
The bulk of the data~\cite{h1-fps1,h1-fps2,zeus-lrglps} was taken at electron and proton beam energies of 
$E_e\simeq27.5$~GeV and $E_p=920$~GeV, 
respectively, corresponding to an $ep$ centre-of-mass energy of
$\sqrt{s} = 318$~GeV. The earlier ZEUS LPS data~\cite{zeus-lps1} collected at $E_p=820$~GeV are corrected   
to a common $\sqrt{s} = 318$~GeV by using the extrapolation procedure described in section~\ref{sec-grid}. 
The three-fold differential reduced cross sections, $\sigma_r^{D(3)}$($\beta$,~$Q^2$,~$\xip$), are combined.
For the original measurements, the main H1 and ZEUS detectors are used 
to reconstruct $Q^2, W$ and $x$, whereas $M_X$, $\beta$, $\xip$ and $t$ are 
derived from the proton spectrometer measurement 
or from combined information of the proton spectrometers and the main 
detectors.
In table~\ref{tab-data} the data sets used for the combination are listed together with their kinematic ranges and  
integrated luminosities.

\begin{table}
\begin{center}
\begin{footnotesize}
\renewcommand{\arraystretch}{1.2}
\begin{tabular}{|c|c|c|c|c|c|c|c|}
\hline
 Data Set & $Q^2$ range & $\xip$ range & $y$ range & $\beta$ range & $t$ range & Luminosity & Ref.\\
 & [GeV$^2$] & & & & [GeV$^2$] & [pb$^{-1}$] & \\
\hline     
 H1 FPS HERA II & $4 - 700$ & $< 0.1$ & $0.03 - 0.8$ & $0.001 - 1$ & $0.1 - 0.7$ & $156.6$ & ~\cite{h1-fps2}\\                                  
\hline 
H1 FPS HERA I & $2 - 50$ & $< 0.1$ & $0.02 - 0.6$ & $0.004 - 1$ & $0.08 - 0.5$ & $28.4$ & ~\cite{h1-fps1}\\                                  
\hline\hline
 & & & $W$ range & $M_X$ range & & & \\
 & & & [GeV] & [GeV] & & & \\\hline
ZEUS LPS 2 & $2.5 - 120$ & $0.0002 - 0.1$ & $40 - 240$ & $2 - 40$ & $0.09 - 0.55$ & $32.6$ &~\cite{zeus-lrglps} \\
\hline 
ZEUS LPS 1 & $2 - 100$ & $<0.1$ & $25 - 240$ & $>1.5$ & $0.075 - 0.35$ & $3.6$ &~\cite{zeus-lps1}\\  
\hline
\end{tabular}
\end{footnotesize}
\end{center}
\vspace{-0.3cm}

\caption{H1 and ZEUS data sets used for the combination.}
\label{tab-data}
\end{table}

\subsubsection{Restricted {\boldmath $t$} range}
\label{sec-t}

In the individual analyses~\cite{h1-fps1,h1-fps2,zeus-lrglps,zeus-lps1} the 
reduced cross sections are directly measured for ranges of the squared 
four-momentum 
transfer $t$ visible to the proton spectrometers (see table~\ref{tab-data}) 
and extrapolated to the range\footnote{The smallest kinematically accessible value of 
$|t|$ is denoted as $|t_{min}|$.} $|t_{min}| < |t| < 1$ GeV$^2$ (denoted in the
following as `the full $t$ range'), assuming an exponential 
$t$ dependence of the diffractive cross section 
and using the exponential slope measured from the data. 
Due to the uncertainties of the slope parameters measured by H1~\cite{h1-fps2, h1-fps1} and 
ZEUS~\cite{zeus-lrglps,zeus-lps1}, 
this extrapolation 
introduces an additional uncertainty in the normalisation of the cross section. 
To reduce this source of
systematic uncertainty, the H1 and ZEUS cross sections are 
combined in the restricted $t$ range $0.09 < |t| < 0.55$ GeV$^2$ 
covered by the proton spectrometer acceptances of both detectors for the bulk of the data. 
 The correction factors from the visible $t$ range of the `FPS HERA I' and `LPS 1' data samples to the 
restricted $t$ 
range are evaluated by
using the  $t$ dependencies as a function of $x_{\pom}$ measured for each sample.  
The correction factors for the most precise `FPS HERA II' data are applied in bins of $\beta, Q^2$ and $\xpom$.
For the `LPS 2' sample 
the restricted range coincides with the visible range. 
Because of the uncertainty on the exponential
 slope parameter, such factors introduce uncertainties of
2.2\%, 1.1\% and 5\% on the `FPS HERA II', `FPS HERA I' and `LPS 1' data, 
respectively, which are included in the normalisation uncertainty on each 
sample. 
The total normalisation uncertainties of the data samples are listed in table~\ref{tab-t}.
In the restricted $t$ range, these uncertainties 
are in general smaller 
and the average normalisations are in better agreement than in the full $t$ range; 
the ratio of the `FPS HERA II' to the `LPS 2' data averaged over the measured 
data points, which is 
$0.85 \pm 0.01$(stat) $\pm 0.03 $(sys) $^{+0.09}_{-0.12}$(norm)~ 
in the full $t$ range~\cite{h1-fps2}, becomes 
$0.91 \pm 0.01$ (stat) $\pm 0.03$ (sys) $\pm 0.08$ (norm)~
in the restricted $t$ range. Within the uncertainties, the ratio does not show any 
significant $\beta$, $Q^2$ or $\xip$ dependence. 


\begin{table}
\begin{center}
\renewcommand{\arraystretch}{1.2}
\begin{tabular}{|c|c|c|}
\hline
 Data Set & $|t_{min}| < |t| < 1$ GeV$^2$ & $0.09 < |t| < 0.55$ GeV$^2$ \\
\hline
FPS HERA II & $\pm 6 \%$ & $\pm 5\%$ \\
FPS HERA I & $\pm 10 \%$ & $\pm 10 \%$\\
LPS 2 & $+11\%$, $-7\%$ & $\pm 7 \%$\\
LPS 1    & $+12\%$, $-10\%$& $\pm 11 \%$\\ 
\hline
\end{tabular}
\vspace{-0.3cm}

\end{center}
\caption{Normalisation uncertainties in 
the full range $|t| < 1$ GeV$^2$ and in the restricted $t$ range for the data used for the combination.}
\label{tab-t}
\end{table}

\subsubsection{Extrapolation to a common ({\boldmath $Q^2$, $\xip$, $\beta$}) grid}
\label{sec-grid}

The original binning schemes of the $\sigma_r^{D(3)}$
measurements are very different.
In the H1 case the measurements are extracted at fixed $\beta$, whereas for ZEUS the cross section is measured 
at fixed $M_X$; also the $Q^2$~and 
$\xip$~central values differ. Therefore, prior to the combination, the H1 and 
ZEUS data are transformed to a common grid of ($\beta, Q^2, x_{\pom}$)
points.
The grid points are based on the original binning scheme 
of the `FPS HERA II' data. 
The ($Q^2, \xip$) grid points at the lowest $Q^2$ value 
of $2.5$~GeV$^2$ and at the lowest and highest $ x_{\pom}$ values, which are beyond the `FPS HERA II' data grid, are 
taken from the `LPS 2' measurement.

The transformation of a measurement from the original $i^{th}$ point 
($\beta_i, Q^2_i, \xip_i$)  
to the nearest grid point ($\beta_{grid}, Q^2_{grid}, \xip_{grid}$) is performed by 
multiplying the measured cross section by the ratio $\sigma_r^{D(3)}$($\beta_{grid}, Q^2_{grid}, \xip_{grid}$)/ $\sigma_r^{D(3)}$($\beta_{i}, Q^2_{i}, \xip_{i}$) 
calculated with the Next-to-Leading-Order (NLO) DPDF `ZEUS SJ' parameterisation~\cite{zeus-qcdfits}. 
Most of the corrections are smaller than $10$\%, while a few points undergo 
corrections up to $\sim~30$\%. The procedure is checked by using the  
NLO DPDF `H1 Fit B' parameterisation~\cite{h1-lrg1}. 
The resulting difference is treated as a procedural uncertainty on the
combined cross section, as discussed in Section~\ref{sec-procedurals}.

\begin{figure}[hb]
\begin{center}
\includegraphics[scale=0.8]{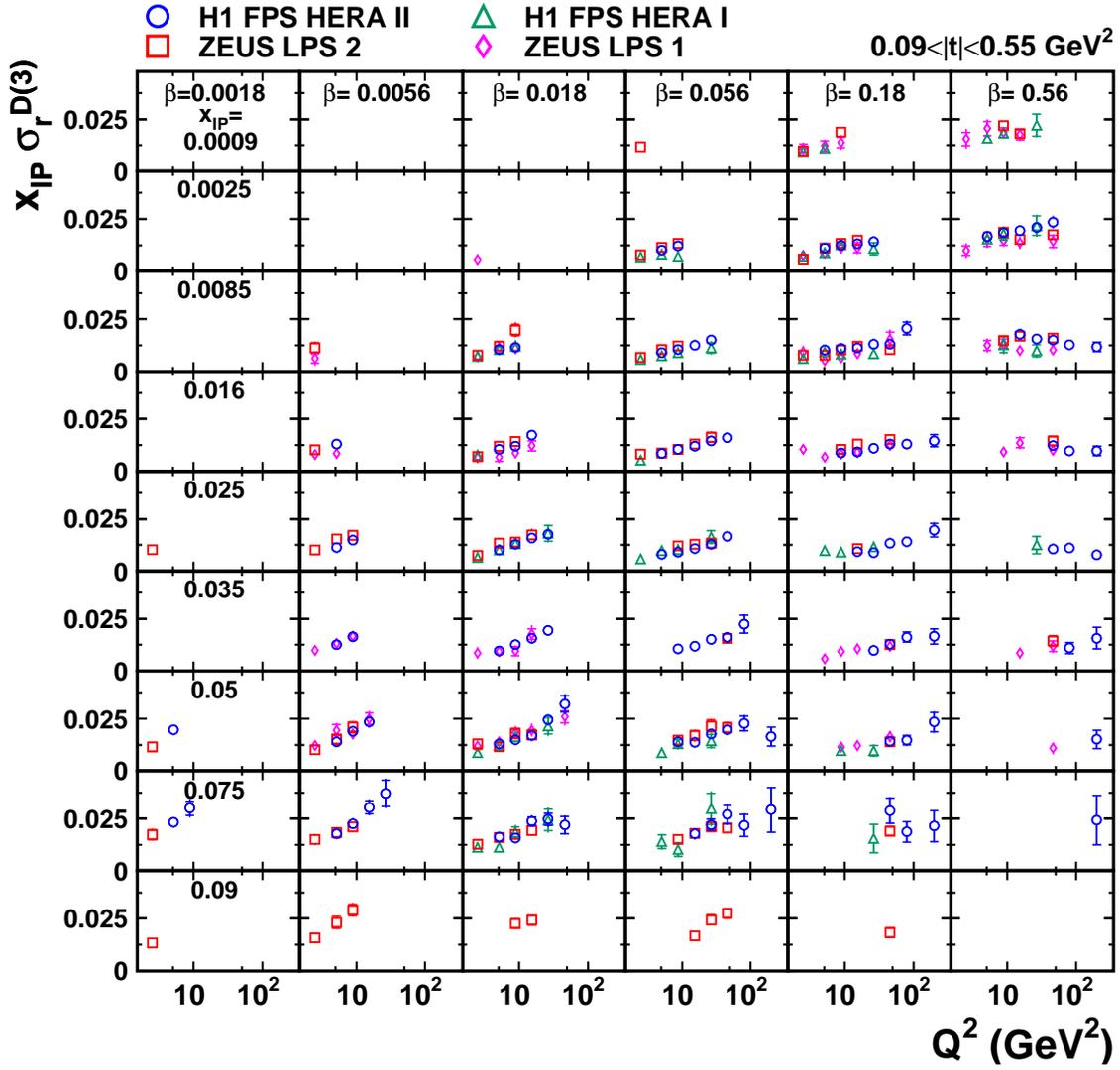}
\caption{Reduced diffractive  cross section $x_{\pom} \, \sigma_r^{D(3)}(\beta,Q^2,x_{\pom})$
for $0.09 < |t| < 0.55 \ {\rm GeV^2}$
as a function of $Q^2$ for different values of $\beta$ and $\xip$.
 The H1 `FPS~HERA II'~\cite{h1-fps2},  H1 `FPS~HERA I'~\cite{h1-fps1}, ZEUS `LPS~2'~\cite{zeus-lrglps} and ZEUS 
 `LPS~1'~\cite{zeus-lps1} data are presented. The inner error bars indicate the statistical uncertainties, the outer 
 error bars show the statistical and systematic
 uncertainties  added in quadrature. Normalisation uncertainties are not included in the error bars of the individual 
measurements. }
\label{fig-q2dep-allinputs}
\end{center}
\end{figure}

The cross sections from all the data sets are shown in figure~\ref{fig-q2dep-allinputs} after correcting to $0.09 < |t| < 0.55$ GeV$^2$ and transforming to the 
common 
grid. 

\subsection{Combination method}
\label{sec-comb}
The combination is based on the $\chisq$~minimisation method
described in~\cite{glazov} and used for previous combined HERA 
results~\cite{:2009wt}. The averaging procedure is based on 
the assumption that at a given kinematic
 point the H1 and ZEUS experiments are 
measuring the same cross section.
~The correlated systematic uncertainties are floated coherently.
~The procedure allows a model independent check of the data consistency and 
leads to a significant reduction of the correlated uncertainties. 

 For an individual data set, the $\chisq$~function is defined as:
\begin{equation}
\chi^2_{exp} (\pmb{m},\pmb{b}) = \sum_i 
\frac{\left[ m^i - \sum_j \gamma^i_j m^i b_j - \mu^i \right]^2}
{\delta^2_{i,stat} \mu^i \left( m^i - \sum_j \gamma^i_j m^i b_j \right) 
+ \left( \delta_{i,uncor} m^i \right)^2} + \sum_j b^2_j~.
\label{eq-chi}
\end{equation}

\noindent
Here $\mu^i$ is the measured cross section value at a point $i$ 
($\beta_i$,~$Q^2_i$,~$\xip_i$), and $\gamma^i_j$, 
$\delta_{i,stat}$ and $\delta_{i,uncor}$ are the relative correlated systematic, 
relative statistical and relative uncorrelated systematic uncertainties, 
respectively. The vector $\pmb{m}$ of quantities $m^i$ expresses the values 
of the combined cross section for each point $i$ and the vector 
$\pmb{b}$ of quantities $b_j$ expresses the shifts 
of the correlated systematic uncertainty sources, $j$, in units of the standard deviation. 
The relative uncertainties $\gamma^i_j$ and 
$\delta_{i,uncor}$ are multiplied by the 
combined cross section $m^i$ in order to take into account the fact 
that the correlated and uncorrelated systematic uncertainties are to a good 
approximation proportional to the central values (multiplicative uncertainties). 
On the other hand, the statistical uncertainties scale with the square root of the 
expected number of events, which is determined by the expected cross 
section, corrected for the biases due to the correlated 
systematic uncertainties. This is taken into account by the 
$\delta^2_{i,stat} \mu^i ( m^i - \sum_j \gamma^i_j m^i b_j )$ 
term. 



If several analyses provide measurements at the same ($\beta$,~$Q^2$, $\xip$) 
values, 
a $\chi^2_{tot}$ is built~\cite{Aaron:2009bp} from the sum of the $\chi^2_{exp}$ of each data set, assuming  the 
individual data sets to be statistically uncorrelated. The $\chi^2_{tot}$ is minimised with respect to the $m^i$ and 
$b_j$ from each data set 
with an iterative procedure.
~The ratio $\chi^2_{min} / n_{dof}$ is a measure of the 
consistency of the data sets.~The number of degrees of freedom, $n_{dof}$, is
calculated as the difference between the total number of measurements and the 
 number of averaged points. The uncertainties of the combined cross sections are evaluated from the 
 $\chi^2_{min}+1$ 
criteria~\cite{glazov,Aaron:2009bp,:2009wt}. 
For some of the ($\beta, Q^2, x_{\pom}$) points there is only one measurement;
however, because of the systematic 
uncertainty correlations such measurements may 
be shifted with 
respect to the original values, and the uncertainties may be reduced.



\subsection{Uncertainties}

\subsubsection{Experimental systematic uncertainties and their correlations}
\label{sec-corrsys}

The input cross sections are published with their statistical and 
 systematic uncertainties. The statistical uncertainties correspond to $\delta_{i,stat}$ in Eq.~(\ref{eq-chi}). The systematic 
 uncertainties are classified as point-to-point correlated or point-to-point uncorrelated, corresponding to $\gamma^i_j$ and 
 $\delta_{i,uncor}$ respectively, according to the information provided in the corresponding publications, as follows: 
\begin{itemize}
\item
 for the two older analyses, `FPS HERA I' and `LPS 1', only the total systematic 
 uncertainties are given~\cite{h1-fps1,zeus-lps1}, with no information on the single contributions and 
 point-to-point 
 correlations. For these two samples only the normalisation uncertainties~(table~\ref{tab-t}) are considered among the correlated 
systematics, while the remaining uncertainties are treated as uncorrelated;
\item
 for the sample `FPS HERA II' all the systematic sources discussed in~\cite{h1-fps2} are treated as point-to-point correlated. 
 The hadronic energy scale uncertainty is taken as correlated separately for 
 $\xip < 0.012$ and $\xip > 0.012$. This is to account for the different sensitivity to this 
 systematic source for the two $\xip$ regions, where different methods are
 used to reconstruct the variable $\beta$, which are typically sensitive to different  
 regions of the H1 central  calorimeter. 
 For $\xpom<0.012$, where 
 the mass  $M_X$ of the hadronic final state is used to reconstruct $\beta$, the 
 effect on the cross section due to the hadronic energy scale uncertainty is 4\% on average and reaches 
 6.7\%. For $\xpom>0.012$, where $\beta$ is reconstructed with the 
 leading proton energy measured by the FPS, the cross section shows almost 
 no sensitivity to the hadronic energy scale;
\item
 in the `LPS 2' case, the total systematic uncertainties quoted in~\cite{zeus-lrglps} are 
decomposed in
 correlated and uncorrelated 
following the prescriptions in~\cite{zeus-qcdfits}. They are symmetrised
 by taking the average of the positive and negative uncertainties. 
\end{itemize}

In the H1 `FPS HERA II' analysis, the systematic effects related to the leading proton measurement are considered 
as correlated and derived from the variation of the acceptance in the $\xpom$ and $t$ bins when 
shifting
the FPS energy scale 
and transverse momentum within the estimated uncertainties~\cite{h1-fps2}.   
In the ZEUS 'LPS 2' case, the systematic uncertainty
related to the leading proton measurement is dominated by the
incomplete knowledge of the beam optics, of the position of the beamline
aperture limitations and of the intrinsic transverse-momentum spread of the proton beam at the interaction point. 
The beam optics contribution is largely independent of the kinematic variables and therefore is taken as a
normalisation uncertainty~\cite{zeus-lrglps}.
The other contributions are quantified by
varying the cut on the distance of closest approach of the reconstructed proton track to the beampipe, and the value of the intrinsic transverse-momentum spread assumed in the simulation. They are treated as uncorrelated uncertainties.


All the H1 systematic uncertainties are treated as independent of the ZEUS uncertainties, and vice versa. 
Possible effects due to correlations between the two experiments are 
taken into account in the procedural uncertainties, 
discussed in Section~\ref{sec-procedurals}.
In total, 23 independent sources of correlated systematic
uncertainties are considered, including the global normalisation for each sample. The full list is given in 
table~\ref{tab-sys}.

\subsubsection{Procedural uncertainties}
\label{sec-procedurals}

The following uncertainties on the combined cross sections due to the combination procedure 
 are studied:
\begin{itemize}
\item
The $\chi^2$ function given by Eq.~(\ref{eq-chi}) treats all 
systematic uncertainties as multiplicative, i.e. proportional 
to the expected central values. While this generally holds for 
the normalisation uncertainties, it may not be the case for the 
other uncertainties.~To study the sensitivity of the average 
result to this issue, an alternative averaging is performed. 
Only the normalisation uncertainty and those related to the $t$ 
reconstruction 
(the uncertainties on the `FPS HERA II' proton $p_x$, $p_y$ reconstruction and on the `FPS HERA II' and `LPS 2' $t$ 
reweighting) 
which, for the reasons explained in Section~\ref{sec-t}, 
can affect the normalisation, are taken as multiplicative,  
while all other uncertainties are treated as additive. The difference 
between this average and the nominal result is of the order of 1\% 
on average and 6.4\% at most.
\item
The H1 and ZEUS experiments use similar methods for detector calibration, apply similar 
reweighting to the Monte Carlo models used for the acceptance corrections and employ similar Monte Carlo models for QED 
radiative corrections, 
for the hadronic final state simulation and for background subtraction. 
Such similarities may lead to 
correlations between the measurements of the two experiments. 
Three systematic source are identified as the most likely to be correlated between the two experiments.
These are 
the electromagnetic energy scale 
and the reweighting of the 
simulation in $\xip$ and $t$. Averages are formed for each of the $2^3$ possible
assumptions on 
the presence of correlations of these systematic uncertainties 
between the experiments and are compared with the nominal average for which all sources are assumed to be 
uncorrelated. 
 The maximum difference
      between the nominal and the alternative averages is
      taken as an uncertainty.
It is 1.4\% on average and 6.6\% at most, 
with no particular dependence on the kinematics.

\item
The bias introduced by transforming the data to the common grid
 (see Section~\ref{sec-grid}) is studied by using 
correction factors obtained from the NLO DPDF `H1 Fit B'~\cite{h1-lrg1} 
parameterisation. 
For a few bins this changes the result by 
up to 8\%, but 
the average effect is 1.2\%.

 \item The averaging procedure shifts the H1 hadronic energy scale at $\xip < 0.012$ by substantially more than $1\sigma$ 
of the nominal value 
(see Section~\ref{sec-res}). To study the 
sensitivity of the average result to the treatment of the uncertainty on the H1 
hadronic energy scale, an alternative averaging is performed for which this uncertainty is considered as 
point-to-point uncorrelated. The difference between the alternative and nominal results is 0.9\% on average and 
reaches 
8.7\% at low 
$\xpom$. 
\end{itemize}

For each combined data point the difference between the average obtained by considering 
each of the procedural effects and the nominal average is calculated and summed 
in quadrature. The effect of the procedural 
uncertainties is 2.9\% on average and 9.3\% at most. 

\section{Results}
\label{sec-res}

\begin{figure}[hb]
\begin{center}
\includegraphics[scale=0.8]{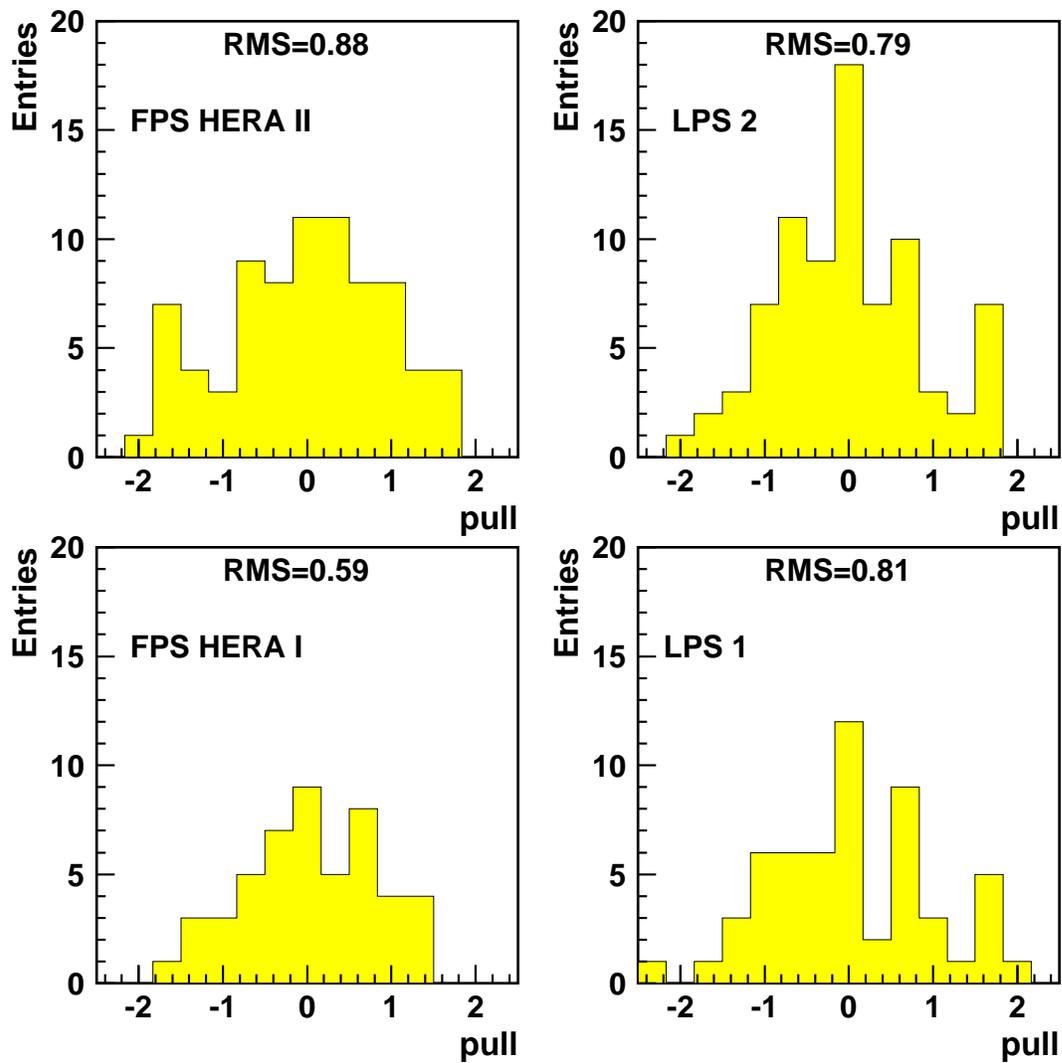}
\caption{Pull distributions 
for the individual data sets.
 The root mean square gives the root mean square of the distributions.
}
\label{fig-pulls}
\end{center}
\end{figure}

In the minimisation procedure, 352 data 
points are combined to 191 cross section measurements.~The data show good 
consistency, with 
$\chi^2_{min} / n_{dof} = 133 / 161$.
The distributions of pulls~\cite{:2009wt}, shown in figure~\ref{fig-pulls} for each data set, 
 exhibit no significant tensions.
 For data with no correlated systematic uncertainties pulls are expected to follow 
 Gaussian distributions with zero mean and unit width. 
 Correlated systematic uncertainties lead to narrowed pull distributions. 

~The effects of the combination on the correlated systematic 
uncertainties are summarised in table~\ref{tab-sys} in terms of shifts 
in units of the original uncertainty and of values of the final uncertainties 
as percentages of the originals. 
The combined cross section values are given in table~\ref{tab-results}  
together with statistical, uncorrelated systematic, correlated systematic, experimental, procedural   
and total uncertainties. The experimental uncertainty is obtained as the quadratic sum of the statistical, 
uncorrelated 
systematic and correlated systematic uncertainties. The total 
uncertainty is defined as the  quadratic sum of the experimental and procedural uncertainties. The 
full information about correlations 
can be found elsewhere~\cite{website}. As the global 
normalisations of the input data sets are fitted as correlated systematic uncertainties, the normalisation uncertainty on the 
combined data is included in the correlated systematic uncertainty given in table~\ref{tab-results}. 

Most of the 23 correlated systematic uncertainties shift by less than $0.5~\sigma$ of the 
nominal value in the averaging procedure. None of them shifts by substantially more than 
$1 \sigma$, with the exception of the hadronic energy scale for $\xip<0.012$ 
for the `FPS HERA II' sample. 
Detailed studies show that there is a tension between the H1 `FPS HERA II' and ZEUS `LPS 2' data at low $\xip$; the 
average ratio of the H1 to ZEUS cross sections is above $1.0$ for 
$\beta>0.1$ and below $0.9$ for $\beta<0.1$. The H1
cross section 
uncertainty is positively correlated with the hadronic energy scale for $\beta>0.1$ and anti-correlated for $\beta<0.1$. As
a result, the combination shifts the H1 cross sections for $\xip<0.012$ in the direction opposite to the cross section uncertainty due to the 
H1 hadronic energy scale. Conversely the combined statistical and uncorrelated uncertainty on the ZEUS data is much 
larger than the 
ZEUS hadronic energy scale uncertainty; consequently 
the fit is less sensitive to the ZEUS hadronic energy scale.

\begin{figure}[hb]
\begin{center}
\includegraphics[scale=0.8]{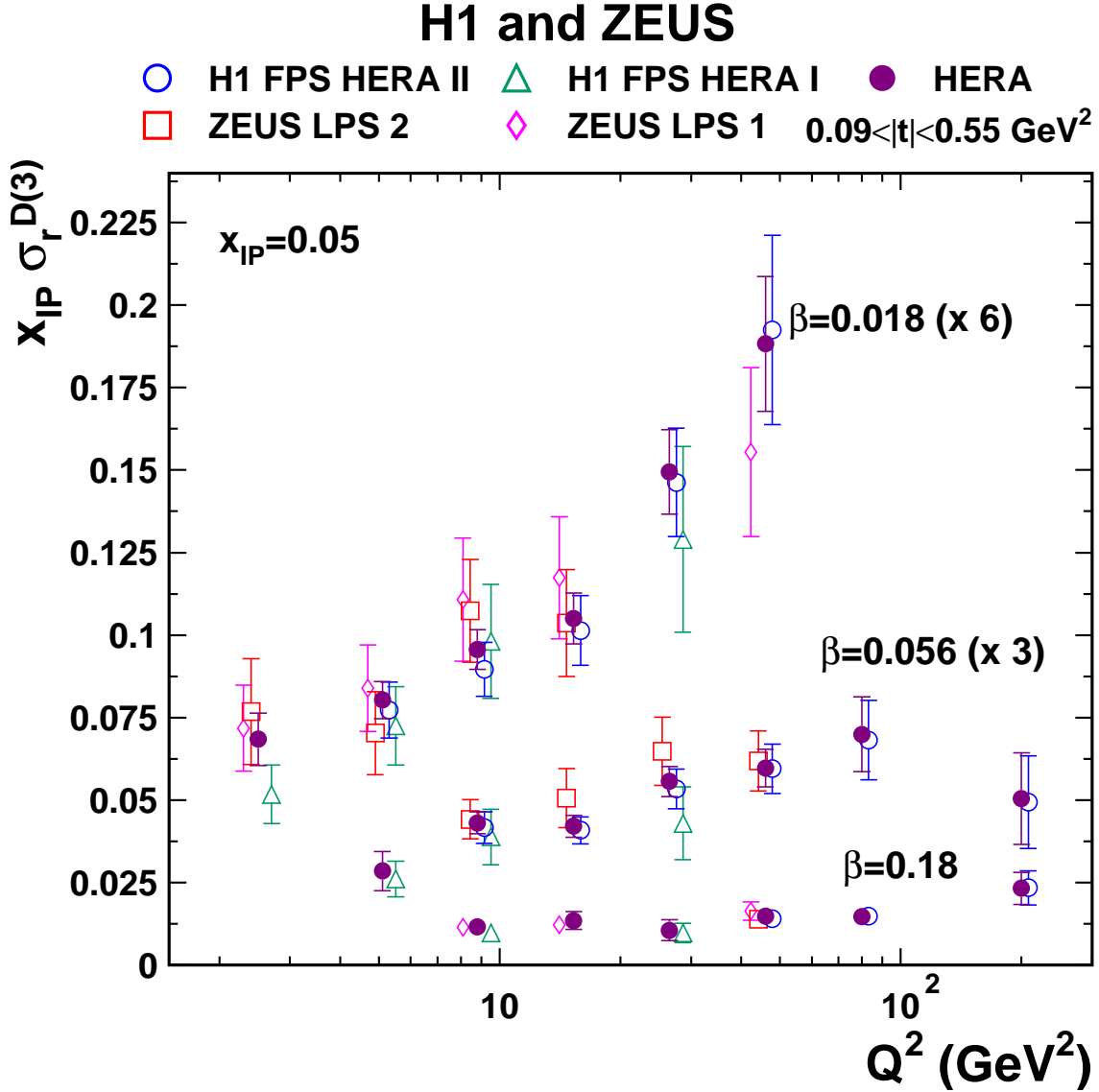}
\caption{Reduced diffractive  cross section $x_{\pom} \, \sigma_r^{D(3)}(\beta,Q^2,x_{\pom})$ for $0.09 < |t| < 0.55 \ {\rm GeV^2}$
as a function of $Q^2$ for different values of $\beta$ at $\xip = 0.05$. 
 The combined data are compared to the H1 and ZEUS data input to the averaging procedure.
 The error bars indicate 
 the statistical and systematic uncertainties  added in quadrature for the input measurements and the statistical, 
systematic and 
 procedural uncertainties added in quadrature for the combined points. Normalisation uncertainties are not included in 
the error bars of 
 the individual measurements, whereas they are included in the error bars of the combined points.}
\label{fig-q2sel-all}
\end{center}
\end{figure}

\begin{figure}[hb]
\begin{center}
\includegraphics[scale=0.8]{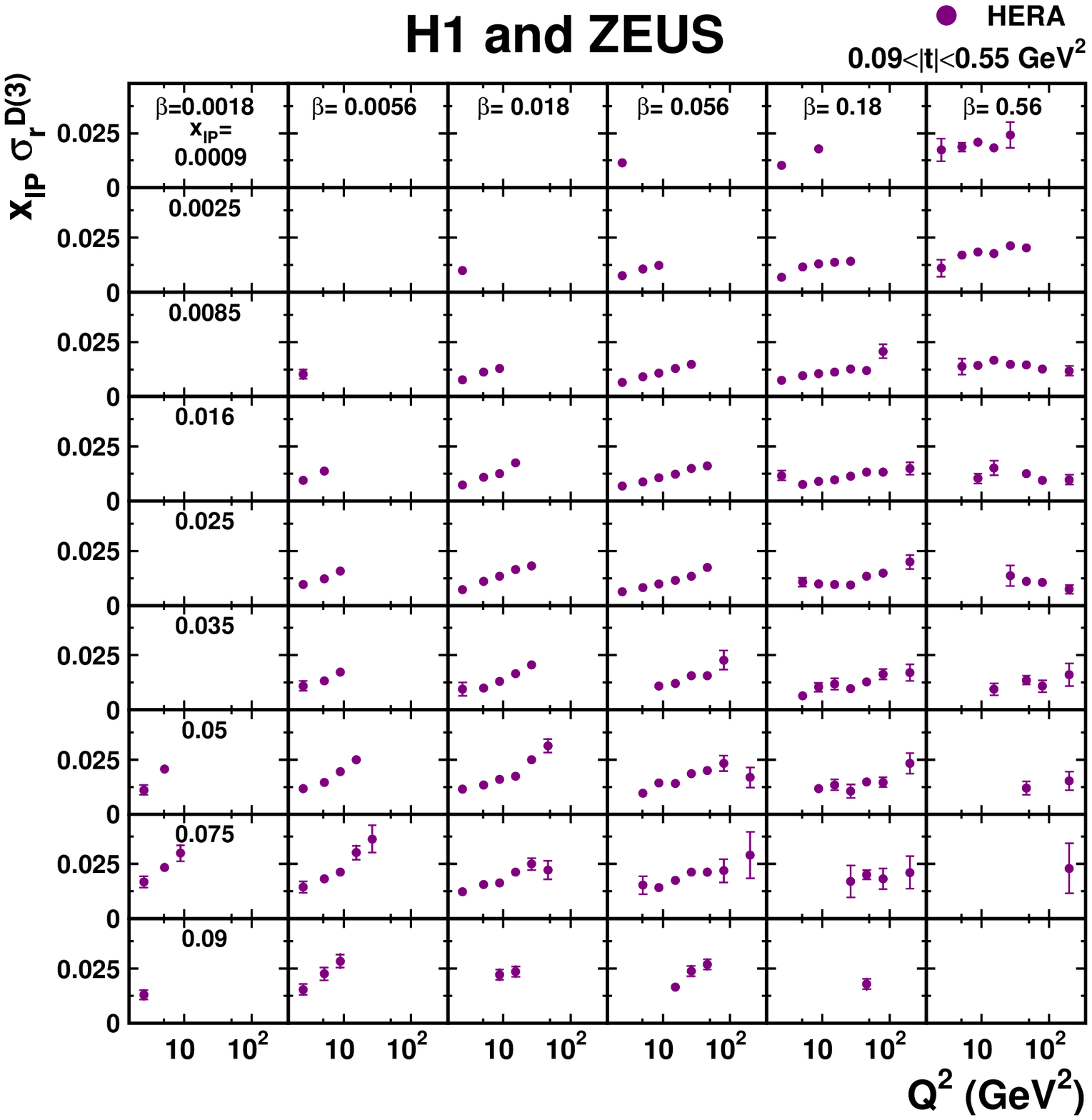}
\caption{HERA combined reduced diffractive cross section 
 $x_{\pom} \, \sigma_r^{D(3)}(\beta,Q^2,x_{\pom})$ for $0.09 < |t| < 0.55 \ {\rm GeV^2}$ as a function of $Q^2$ for 
 different values of $\beta$ and $\xip$. 
 The error bars indicate the statistical, systematic and procedural uncertainties added in quadrature. The normalisation 
uncertainty is 
 included.}
\label{fig-q2dep-comb}
\end{center}
\end{figure}

\begin{figure}[hb]
\begin{center}
\includegraphics[scale=0.8]{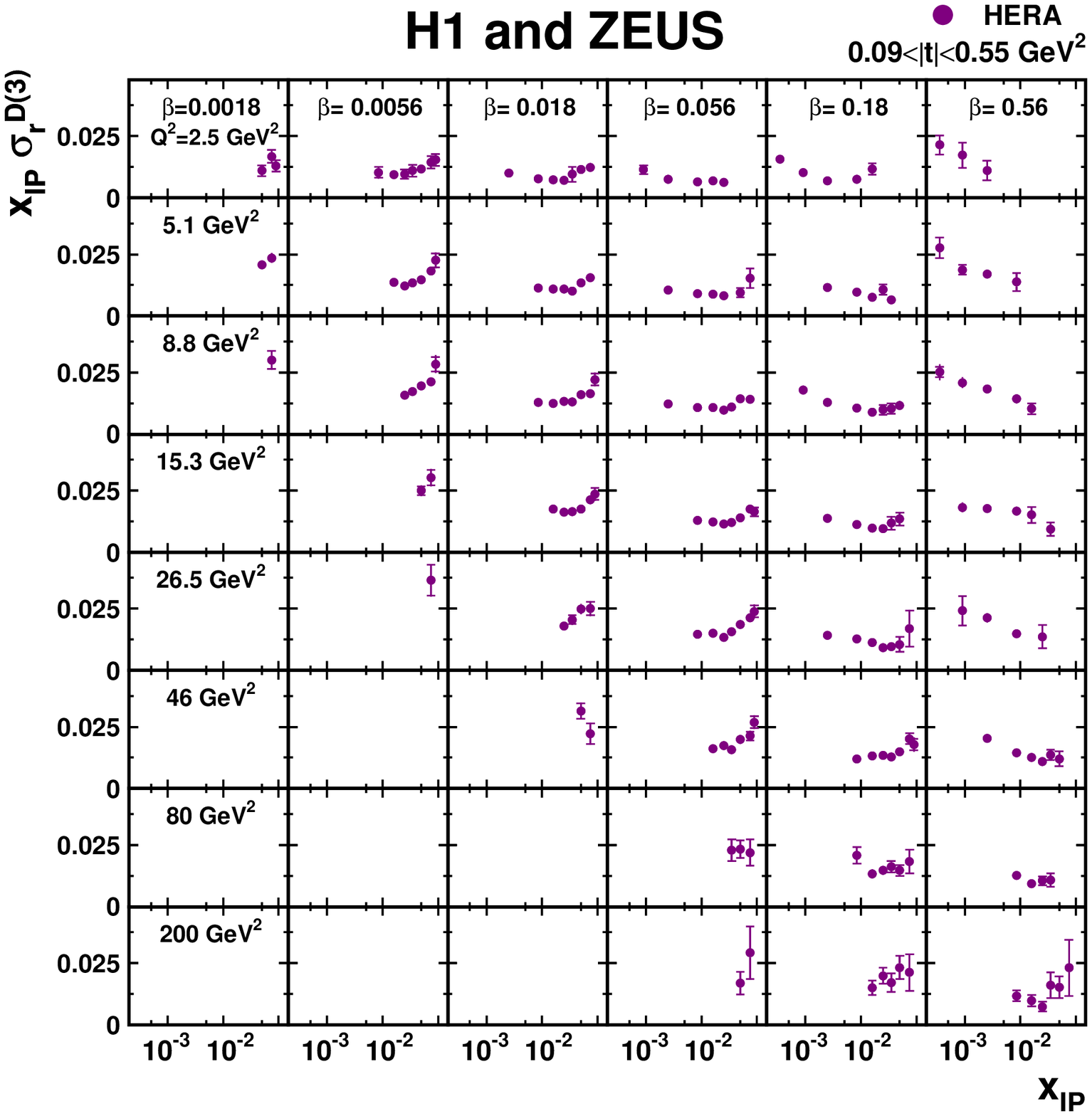}
\caption{HERA combined reduced diffractive cross section 
 $x_{\pom} \, \sigma_r^{D(3)}(\beta,Q^2,x_{\pom})$ for $0.09 < |t| < 0.55 \ {\rm GeV^2}$ as a function of $\xip$ for 
 different values of $\beta$ and $Q^2$. The error bars indicate the statistical, systematic and procedural uncertainties 
added in 
 quadrature. The normalisation uncertainty is included.}
\label{fig-xpdep-comb}
\end{center}
\end{figure}

The influence of several correlated systematic uncertainties is reduced 
significantly for the combined result. Specifically, the uncertainty on the 
FPS proton energy measurement and the normalisation uncertainties on the 
`FPS HERA I' and `LPS 1' samples are reduced by more than a factor of 2.
The H1 hadronic energy scale uncertainty for the low $\xip$-range ($\xip<0.012$) and the 
ZEUS hadronic energy scale uncertainty are reduced to around 55\% of those for the individual data sets. 
Since H1 and ZEUS use different 
reconstruction methods, similar systematic sources influence the measured 
 cross section differently. 
Therefore, requiring the cross sections to be consistent at all 
($\beta$, $Q^2$, $\xip$) points constrains the systematic uncertainties efficiently.
Due to this cross calibration effect,
the combined measurement shows an average improvement of the experimental uncertainty of about 27\% with respect 
to the most precise single data set, 
`FPS HERA II', though the latter data set contains five times more events than
the second largest data set, `LPS 2'. 
The correlated  part of the experimental uncertainty is reduced from about 69\% in~\cite{h1-fps2} to 49\% in the 
combined 
measurement. The statistical, experimental and procedural uncertainties on the combined data are  on average 11\%, 13.8\% and 
2.9\%, respectively. The total uncertainty on the cross section is 14.3\% on average and is 6\% for the most precise
 points. The normalisation uncertainty, which contributes to 
the correlated systematic uncertainty on the combined data, 
is on average 4\%.
The combined result extends the kinematic 
coverage with respect to the H1 and ZEUS measurements taken separately 
and the resulting
cross section covers the region $2.5 \leq Q^2 \leq 200$~GeV$^2$, $0.0018 \leq \beta \leq 0.816$
and $0.00035 \leq x_{\pom} \leq 0.09$, for $0.09 < |t| < 0.55$~GeV$^2$.
Figure~\ref{fig-q2sel-all} shows the combined cross section as a function of $Q^2$~at
 $x_{\pom} = 0.05$, for different values of $\beta$, compared with the individual measurements used for the 
combination.~The reduction of the 
total uncertainty
of the HERA measurement compared to the input cross sections is visible.
~The derivative of the reduced cross section as a function of $\log(Q^2)$ decreases with 
$\beta$,
a feature characteristic of the scaling violations in
diffractive DIS, which are now measured precisely from proton-tagged as well
as LRG data.
Figures~\ref{fig-q2dep-comb} and~\ref{fig-xpdep-comb} show the HERA combined diffractive reduced cross sections as a 
function of $Q^2$ 
and $\xip$, respectively. 

At low $\xip \lesssim 0.01$, 
where the proton spectrometer data are free from proton dissociation contributions, the combined 
data provide the most precise determination of the 
absolute normalisation of the diffractive cross section. 

\section{Conclusions}

The reduced diffractive cross sections, $\sigma_r^{D(3)}(ep \to eXp)$, measured by the H1 and 
ZEUS Collaborations by using proton spectrometers to detect the leading protons  
are combined. The input data 
 from the two experiments are consistent with a $\chi^2_{min} / n_{dof} = 133 / 161$. 
 The combination of the measurements results in more precise and kinematically extended diffractive DIS data in the 
$t$-range $0.09 < |t| < 0.55$~GeV$^2$. The total uncertainty on the cross section measurement is 6\% for the most 
 precise  points.
The combined data provide the most precise determination of the absolute normalisation of the $ep\rightarrow eXp$ cross section.



\section*{Acknowledgements}

We are grateful to the HERA machine group whose outstanding
efforts have made these experiments possible.
We appreciate the contributions to the construction and maintenance of the H1 and ZEUS detectors of many people who are not listed as authors.
We thank our funding agencies for financial 
support, the DESY technical staff for continuous assistance and the 
DESY directorate for their support and for the hospitality they extended to the non-DESY members of the collaborations. 



\newpage

\begin{table}[!ht]
\begin{center}
\begin{normalsize}
\begin{tabular}{|l|r|c|} 
\hline
Source & Shift ($\sigma$ units) & Reduction factor \% \\ 
\hline
FPS HERA II hadronic energy scale $\xip<0.012$  & $-1.61~~~~$ & $56.9$      \\
FPS HERA II hadronic energy scale $\xip>0.012$  & $~0.13~~~~$ & $99.8$      \\
FPS HERA II electromagnetic energy scale        & $~0.49~~~~$ & $85.9$      \\
FPS HERA II electron angle                      & $~0.67~~~~$ & $66.6$   \\
FPS HERA II $\beta$~reweighting                 & $~0.15~~~~$ & $90.4$   \\
FPS HERA II $\xip$~reweighting                  & $~0.05~~~~$ & $98.3$ \\
FPS HERA II $t$~reweighting                     & $~0.70~~~~$ & $79.8$        \\
FPS HERA II $Q^2$~reweighting                   & $~0.09~~~~$ & $97.6$     \\
FPS HERA II proton energy                       & $~0.05~~~~$ & $45.6$     \\
FPS HERA II proton $p_x$                        & $~0.62~~~~$ & $74.5$    \\
FPS HERA II proton $p_y$                        & $~0.27~~~~$ & $86.5$   \\
FPS HERA II vertex reconstruction               & $~0.07~~~~$ & $97.0$   \\
FPS HERA II background subtraction              & $~0.84~~~~$ & $89.9$   \\
FPS HERA II bin centre corrections              & $-1.05~~~~$ & $87.3$   \\
FPS HERA II global normalisation                & $-0.39~~~~$ & $84.4$     \\
FPS HERA I global normalisation                 & $~0.81~~~~$ & $48.9$     \\
\hline
LPS 2 hadronic energy scale                     & $-0.02~~~~$ & $55.0$     \\
LPS 2 electromagnetic energy scale              & $-0.14~~~~$ & $62.4$   \\
LPS 2 $\xip$~reweighting                        & $-0.32~~~~$ & $98.2$ \\
LPS 2 $t$~reweighting                           & $-0.26~~~~$ & $86.4$       \\
LPS 2 background subtraction                    & $~0.40~~~~$ & $94.9$     \\
LPS 2 global normalisation                      & $-0.53~~~~$ & $67.7$  \\
LPS 1 global normalisation                      & $~0.86~~~~$ & $44.1$   \\
\hline
\end{tabular}
\end{normalsize}
\caption{Sources of point-to-point correlated systematic uncertainties 
considered in the combination. For each source the shifts resulting 
from the combination in units of the original uncertainty 
and the values of the final uncertainties as percentages of the original 
are given.}
\label{tab-sys}
\end{center}
\end{table}

\begin{table}[ht!]
\begin{center}
\begin{tabular}{|c|c|c|c|r@{.}l|r@{.}l|r@{.}l|r@{.}l|r@{.}l|r@{.}l|} 
\hline
$Q^2$ & $\beta$ & $x_{I\!\!P}$ & $x_{I\!\!P} \sigma_r^{D(3)}$ 
& \multicolumn{2}{c|}{$\delta_{stat}$}
& \multicolumn{2}{c|}{$\delta_{uncor}$}
& \multicolumn{2}{c|}{$\delta_{cor}$}
& \multicolumn{2}{c|}{$\delta_{exp}$}
& \multicolumn{2}{c|}{$\delta_{proc}$}
& \multicolumn{2}{c|}{$\delta_{tot}$} \\ 
(GeV$^2$) &  &  &  
& \multicolumn{2}{c|}{(\%)}
& \multicolumn{2}{c|}{(\%)}
& \multicolumn{2}{c|}{(\%)}
& \multicolumn{2}{c|}{(\%)}
& \multicolumn{2}{c|}{(\%)}
& \multicolumn{2}{c|}{(\%)} \\ 
\hline
   2.5 &   0.0018 &   0.0500 &   0.0110 &  19&  &   5&8 &   4&7 &  21&  &  7&6 &  22&  \\
   2.5 &   0.0018 &   0.0750 &   0.0166 &  14&  &   6&9 &   5&3 &  17&  &  7&6 &  18&  \\
   2.5 &   0.0018 &   0.0900 &   0.0128 &  14&  &   9&6 &   5&1 &  18&  &  7&9 &  20&  \\
   2.5 &   0.0056 &   0.0085 &   0.0101 &  19&  &  11&  &   7&6 &  23&  &  9&3 &  25&  \\
   2.5 &   0.0056 &   0.0160 &   0.0093 &  12&  &   6&9 &   5&1 &  14&  &  3&9 &  15&  \\
   2.5 &   0.0056 &   0.0250 &   0.0096 &  16&  &   9&8 &   5&0 &  20&  &  4&6 &  20&  \\
   2.5 &   0.0056 &   0.0350 &   0.0110 &  18&  &  11&  &   4&9 &  22&  &  2&3 &  22&  \\
   2.5 &   0.0056 &   0.0500 &   0.0117 &   9&8 &   6&4 &   5&3 &  13&  &  1&5 &  13&  \\
   2.5 &   0.0056 &   0.0750 &   0.0143 &  14&  &  11&  &   5&7 &  19&  &  4&7 &  19&  \\
   2.5 &   0.0056 &   0.0900 &   0.0154 &  15&  &   6&4 &   5&7 &  17&  &  4&3 &  17&  \\
   2.5 &   0.0178 &   0.0025 &   0.0099 &  14&  &   6&8 &   4&5 &  16&  &  8&2 &  18&  \\
   2.5 &   0.0178 &   0.0085 &   0.0076 &   8&3 &   7&1 &   4&5 &  12&  &  1&7 &  12&  \\
   2.5 &   0.0178 &   0.0160 &   0.0073 &   8&2 &   9&5 &   4&5 &  13&  &  1&4 &  13&  \\
   2.5 &   0.0178 &   0.0250 &   0.0071 &   8&8 &   9&2 &   4&5 &  14&  &  1&4 &  14&  \\
   2.5 &   0.0178 &   0.0350 &   0.0095 &  15&  &  29&  &   4&9 &  33&  &  2&3 &  33&  \\
   2.5 &   0.0178 &   0.0500 &   0.0114 &   7&8 &   7&1 &   4&5 &  11&  &  2&2 &  12&  \\
   2.5 &   0.0178 &   0.0750 &   0.0123 &  11&  &   7&8 &   4&9 &  14&  &  1&7 &  14&  \\
   2.5 &   0.0562 &   0.0009 &   0.0114 &  13&  &   8&6 &   5&2 &  16&  &  3&4 &  17&  \\
   2.5 &   0.0562 &   0.0025 &   0.0074 &   9&3 &   5&7 &   4&8 &  12&  &  2&8 &  12&  \\
   2.5 &   0.0562 &   0.0085 &   0.0064 &   9&6 &   6&7 &   4&5 &  13&  &  2&3 &  13&  \\
   2.5 &   0.0562 &   0.0160 &   0.0068 &  10&  &  10&  &   4&6 &  15&  &  4&4 &  16&  \\
   2.5 &   0.0562 &   0.0250 &   0.0063 &  14&  &  14&  &   4&9 &  20&  &  1&9 &  20&  \\
   2.5 &   0.1780 &   0.0003 &   0.0156 &   8&8 &   5&4 &   4&7 &  11&  &  2&6 &  12&  \\
   2.5 &   0.1780 &   0.0009 &   0.0102 &   5&9 &   4&3 &   4&4 &   8&5 &  2&2 &   8&8 \\
   2.5 &   0.1780 &   0.0025 &   0.0068 &   8&0 &   6&3 &   4&7 &  11&  &  2&6 &  12&  \\
   2.5 &   0.1780 &   0.0085 &   0.0074 &   9&3 &  10&  &   4&8 &  15&  &  3&9 &  15&  \\
   2.5 &   0.1780 &   0.0160 &   0.0116 &  18&  &   7&5 &   5&0 &  20&  &  2&3 &  20&  \\
   2.5 &   0.5620 &   0.0003 &   0.0214 &  16&  &   8&8 &   5&0 &  19&  &  2&3 &  19&  \\
   2.5 &   0.5620 &   0.0009 &   0.0172 &  19&  &  23&  &   5&0 &  31&  &  2&3 &  31&  \\
   2.5 &   0.5620 &   0.0025 &   0.0110 &  21&  &  28&  &   4&9 &  36&  &  2&3 &  36&  \\
   5.1 &   0.0018 &   0.0500 &   0.0199 &   5&9 &   0&0 &   6&6 &   8&9 &  1&8 &   9&1 \\
   5.1 &   0.0018 &   0.0750 &   0.0232 &   6&7 &   0&0 &   5&1 &   8&4 &  2&1 &   8&7 \\
   5.1 &   0.0056 &   0.0160 &   0.0135 &   3&9 &   0&6 &   5&9 &   7&1 &  2&0 &   7&4 \\
\hline
\end{tabular}
\caption{Combined reduced cross sections 
$\xip \sigma_r^{D(3)}(\beta,Q^2,x_{\pom})$ for diffractive $ep$ scattering, $ep \to e X p$. The values indicated 
by ~$\delta_{stat}$, $\delta_{uncor}$,
$\delta_{cor}$, $\delta_{exp}$, $\delta_{proc}$ and $\delta_{tot}$ represent the
statistical, uncorrelated systematic, correlated systematic, experimental, procedural and 
total uncertainties, respectively.}
\label{tab-results}
\end{center}
\end{table}

\begin{table}[ht!]
\begin{center}
\begin{tabular}{|c|c|c|c|r@{.}l|r@{.}l|r@{.}l|r@{.}l|r@{.}l|r@{.}l|} 
\hline
$Q^2$ & $\beta$ & $x_{I\!\!P}$ & $x_{I\!\!P} \sigma_r^{D(3)}$ 
& \multicolumn{2}{c|}{$\delta_{stat}$}
& \multicolumn{2}{c|}{$\delta_{uncor}$}
& \multicolumn{2}{c|}{$\delta_{cor}$}
& \multicolumn{2}{c|}{$\delta_{exp}$}
& \multicolumn{2}{c|}{$\delta_{proc}$}
& \multicolumn{2}{c|}{$\delta_{tot}$} \\ 
(GeV$^2$) &  &  &  
& \multicolumn{2}{c|}{(\%)}
& \multicolumn{2}{c|}{(\%)}
& \multicolumn{2}{c|}{(\%)}
& \multicolumn{2}{c|}{(\%)}
& \multicolumn{2}{c|}{(\%)}
& \multicolumn{2}{c|}{(\%)} \\ 
\hline
   5.1 &   0.0056 &   0.0250 &   0.0120 &   3&4 &   0&3 &   5&2 &   6&2 &   2&0 &   6&6 \\
   5.1 &   0.0056 &   0.0350 &   0.0134 &   4&0 &   0&6 &   4&7 &   6&2 &   1&5 &   6&3 \\
   5.1 &   0.0056 &   0.0500 &   0.0147 &   3&9 &   0&6 &   5&4 &   6&7 &   3&4 &   7&5 \\
   5.1 &   0.0056 &   0.0750 &   0.0180 &   5&7 &   1&3 &   6&1 &   8&4 &   3&7 &   9&2 \\
   5.1 &   0.0056 &   0.0900 &   0.0224 &  12&  &   3&8 &   4&9 &  14&  &   3&1 &  14&  \\
   5.1 &   0.0178 &   0.0085 &   0.0120 &   2&6 &   0&4 &   5&9 &   6&4 &   7&6 &  10&  \\
   5.1 &   0.0178 &   0.0160 &   0.0111 &   2&6 &   0&2 &   5&2 &   5&8 &   2&8 &   6&5 \\
   5.1 &   0.0178 &   0.0250 &   0.0109 &   3&0 &   0&5 &   5&2 &   6&0 &   2&2 &   6&4 \\
   5.1 &   0.0178 &   0.0350 &   0.0101 &   4&3 &   0&6 &   5&2 &   6&8 &   2&2 &   7&2 \\
   5.1 &   0.0178 &   0.0500 &   0.0134 &   4&1 &   1&4 &   5&1 &   6&7 &   2&2 &   7&0 \\
   5.1 &   0.0178 &   0.0750 &   0.0154 &   6&4 &   2&2 &   4&8 &   8&3 &   2&9 &   8&8 \\
   5.1 &   0.0562 &   0.0025 &   0.0107 &   2&4 &   0&2 &   5&0 &   5&6 &   3&4 &   6&8 \\
   5.1 &   0.0562 &   0.0085 &   0.0088 &   2&7 &   0&3 &   5&0 &   5&7 &   3&5 &   6&7 \\
   5.1 &   0.0562 &   0.0160 &   0.0088 &   3&2 &   0&3 &   5&1 &   6&0 &   2&7 &   6&6 \\
   5.1 &   0.0562 &   0.0250 &   0.0084 &   4&5 &   0&7 &   5&0 &   6&7 &   3&1 &   7&4 \\
   5.1 &   0.0562 &   0.0500 &   0.0095 &  16&  &  13&  &   4&9 &  21&  &   1&9 &  21&  \\
   5.1 &   0.0562 &   0.0750 &   0.0153 &  23&  &  14&  &   5&0 &  27&  &   1&9 &  27&  \\
   5.1 &   0.1780 &   0.0009 &   0.0121 &  11&  &   7&4 &   4&9 &  14&  &  11&  &  18&  \\
   5.1 &   0.1780 &   0.0025 &   0.0118 &   1&6 &   0&2 &   5&9 &   6&1 &   4&2 &   7&4 \\
   5.1 &   0.1780 &   0.0085 &   0.0095 &   2&8 &   0&5 &   5&0 &   5&8 &   3&5 &   6&7 \\
   5.1 &   0.1780 &   0.0160 &   0.0075 &  14&  &  12&  &   4&9 &  19&  &   2&3 &  19&  \\
   5.1 &   0.1780 &   0.0250 &   0.0107 &  13&  &  13&  &   4&9 &  20&  &   1&9 &  20&  \\
   5.1 &   0.1780 &   0.0350 &   0.0065 &  20&  &  14&  &   5&0 &  25&  &   2&3 &  25&  \\
   5.1 &   0.5620 &   0.0003 &   0.0275 &  13&  &   8&2 &   4&9 &  16&  &   2&3 &  16&  \\
   5.1 &   0.5620 &   0.0009 &   0.0187 &   7&0 &   8&0 &   4&6 &  12&  &   1&8 &  12&  \\
   5.1 &   0.5620 &   0.0025 &   0.0153 &   1&4 &   0&1 &   6&1 &   6&2 &   6&1 &   8&7 \\
   5.1 &   0.5620 &   0.0085 &   0.0137 &  19&  &  19&  &   4&9 &  27&  &   2&3 &  27&  \\
   8.8 &   0.0018 &   0.0750 &   0.0288 &  12&  &   0&0 &   6&2 &  13&  &   1&5 &  13&  \\
   8.8 &   0.0056 &   0.0250 &   0.0152 &   5&0 &   0&8 &   5&1 &   7&2 &   2&0 &   7&5 \\
   8.8 &   0.0056 &   0.0350 &   0.0171 &   5&1 &   1&2 &   4&9 &   7&2 &   1&7 &   7&4 \\
   8.8 &   0.0056 &   0.0500 &   0.0197 &   4&1 &   1&2 &   4&6 &   6&3 &   1&6 &   6&5 \\
   8.8 &   0.0056 &   0.0750 &   0.0212 &   5&9 &   1&1 &   4&8 &   7&7 &   3&8 &   8&6 \\
   8.8 &   0.0056 &   0.0900 &   0.0281 &   9&6 &   4&4 &   5&0 &  12&  &   5&7 &  13&  \\
   8.8 &   0.0178 &   0.0085 &   0.0128 &   4&2 &   0&9 &   5&1 &   6&7 &   4&0 &   7&8 \\
   8.8 &   0.0178 &   0.0160 &   0.0124 &   3&1 &   0&6 &   4&9 &   5&8 &   1&5 &   6&0 \\
   8.8 &   0.0178 &   0.0250 &   0.0133 &   3&4 &   0&6 &   4&8 &   5&9 &   1&5 &   6&1 \\
\hline
\end{tabular}
\captcont{continued}
\label{tab:xsec:b}
\end{center}
\end{table}

\begin{table}[ht!]
\begin{center}
\begin{tabular}{|c|c|c|c|r@{.}l|r@{.}l|r@{.}l|r@{.}l|r@{.}l|r@{.}l|} 
\hline
$Q^2$ & $\beta$ & $x_{I\!\!P}$ & $x_{I\!\!P} \sigma_r^{D(3)}$ 
& \multicolumn{2}{c|}{$\delta_{stat}$}
& \multicolumn{2}{c|}{$\delta_{uncor}$}
& \multicolumn{2}{c|}{$\delta_{cor}$}
& \multicolumn{2}{c|}{$\delta_{exp}$}
& \multicolumn{2}{c|}{$\delta_{proc}$}
& \multicolumn{2}{c|}{$\delta_{tot}$} \\ 
(GeV$^2$) &  &  &  
& \multicolumn{2}{c|}{(\%)}
& \multicolumn{2}{c|}{(\%)}
& \multicolumn{2}{c|}{(\%)}
& \multicolumn{2}{c|}{(\%)}
& \multicolumn{2}{c|}{(\%)}
& \multicolumn{2}{c|}{(\%)} \\ 
\hline
   8.8 &   0.0178 &   0.0350 &   0.0130 &   4&5 &   0&5 &   4&8 &   6&6 &   1&4 &   6&8 \\
   8.8 &   0.0178 &   0.0500 &   0.0159 &   3&8 &   1&0 &   4&6 &   6&1 &   1&5 &   6&3 \\
   8.8 &   0.0178 &   0.0750 &   0.0162 &   5&6 &   1&7 &   4&8 &   7&6 &   2&3 &   8&0 \\
   8.8 &   0.0178 &   0.0900 &   0.0220 &   9&5 &   5&9 &   5&0 &  12&  &   2&7 &  13&  \\
   8.8 &   0.0562 &   0.0025 &   0.0125 &   3&4 &   0&4 &   5&0 &   6&1 &   3&8 &   7&1 \\
   8.8 &   0.0562 &   0.0085 &   0.0106 &   3&2 &   0&6 &   5&0 &   6&0 &   2&0 &   6&3 \\
   8.8 &   0.0562 &   0.0160 &   0.0108 &   2&9 &   0&2 &   5&0 &   5&8 &   2&7 &   6&4 \\
   8.8 &   0.0562 &   0.0250 &   0.0098 &   3&6 &   0&5 &   5&0 &   6&2 &   2&5 &   6&7 \\
   8.8 &   0.0562 &   0.0350 &   0.0109 &   5&2 &   0&0 &   4&9 &   7&2 &   2&1 &   7&5 \\
   8.8 &   0.0562 &   0.0500 &   0.0144 &   5&1 &   1&1 &   5&1 &   7&3 &   2&4 &   7&7 \\
   8.8 &   0.0562 &   0.0750 &   0.0140 &  11&  &   4&3 &   4&6 &  12&  &   1&7 &  13&  \\
   8.8 &   0.1780 &   0.0009 &   0.0177 &   7&7 &   2&7 &   5&0 &   9&6 &   1&6 &   9&7 \\
   8.8 &   0.1780 &   0.0025 &   0.0129 &   2&3 &   0&4 &   5&1 &   5&6 &   2&5 &   6&1 \\
   8.8 &   0.1780 &   0.0085 &   0.0104 &   2&6 &   0&4 &   4&6 &   5&3 &   2&7 &   5&9 \\
   8.8 &   0.1780 &   0.0160 &   0.0090 &   3&9 &   0&7 &   5&3 &   6&6 &   2&6 &   7&1 \\
   8.8 &   0.1780 &   0.0250 &   0.0098 &  14&  &  14&  &   4&9 &  21&  &   1&9 &  21&  \\
   8.8 &   0.1780 &   0.0350 &   0.0103 &  17&  &  11&  &   4&9 &  21&  &   2&3 &  21&  \\
   8.8 &   0.1780 &   0.0500 &   0.0116 &  12&  &   8&3 &   4&5 &  15&  &   1&8 &  16&  \\
   8.8 &   0.5620 &   0.0003 &   0.0250 &   7&1 &   4&2 &   4&4 &   9&3 &   8&9 &  13&  \\
   8.8 &   0.5620 &   0.0009 &   0.0207 &   5&6 &   3&5 &   4&4 &   7&9 &   6&7 &  10&  \\
   8.8 &   0.5620 &   0.0025 &   0.0166 &   1&6 &   0&1 &   6&1 &   6&3 &   8&3 &  10&  \\
   8.8 &   0.5620 &   0.0085 &   0.0142 &   8&5 &   4&3 &   4&3 &  10&  &   8&0 &  13&  \\
   8.8 &   0.5620 &   0.0160 &   0.0102 &  17&  &  13&  &   4&4 &  22&  &   2&3 &  22&  \\
  15.3 &   0.0056 &   0.0500 &   0.0245 &   6&7 &   2&2 &   4&9 &   8&6 &   1&1 &   8&7 \\
  15.3 &   0.0056 &   0.0750 &   0.0296 &  10&  &   0&0 &   5&7 &  12&  &   1&6 &  12&  \\
  15.3 &   0.0178 &   0.0160 &   0.0176 &   4&8 &   0&7 &   5&0 &   7&0 &   2&4 &   7&4 \\
  15.3 &   0.0178 &   0.0250 &   0.0164 &   4&4 &   0&7 &   4&8 &   6&6 &   2&4 &   7&0 \\
  15.3 &   0.0178 &   0.0350 &   0.0165 &   5&7 &   1&1 &   4&7 &   7&5 &   1&4 &   7&6 \\
  15.3 &   0.0178 &   0.0500 &   0.0176 &   4&9 &   1&4 &   4&8 &   7&0 &   2&2 &   7&4 \\
  15.3 &   0.0178 &   0.0750 &   0.0211 &   6&7 &   2&1 &   4&8 &   8&5 &   2&6 &   8&9 \\
  15.3 &   0.0178 &   0.0900 &   0.0234 &  10&  &   1&6 &   4&8 &  11&  &   3&3 &  12&  \\
  15.3 &   0.0562 &   0.0085 &   0.0134 &   4&5 &   0&0 &   6&0 &   7&5 &   6&1 &   9&7 \\
  15.3 &   0.0562 &   0.0160 &   0.0122 &   3&9 &   0&3 &   4&9 &   6&3 &   2&5 &   6&8 \\
  15.3 &   0.0562 &   0.0250 &   0.0113 &   4&5 &   0&3 &   4&8 &   6&6 &   1&0 &   6&7 \\
  15.3 &   0.0562 &   0.0350 &   0.0121 &   6&2 &   0&0 &   5&0 &   8&0 &   2&0 &   8&2 \\
  15.3 &   0.0562 &   0.0500 &   0.0140 &   5&7 &   1&1 &   4&9 &   7&6 &   2&0 &   7&8 \\
\hline
\end{tabular}
\captcont{continued}
\label{tab:xsec:c}
\end{center}
\end{table}

\begin{table}[ht!]
\begin{center}
\begin{tabular}{|c|c|c|c|r@{.}l|r@{.}l|r@{.}l|r@{.}l|r@{.}l|r@{.}l|} 
\hline
$Q^2$ & $\beta$ & $x_{I\!\!P}$ & $x_{I\!\!P} \sigma_r^{D(3)}$ 
& \multicolumn{2}{c|}{$\delta_{stat}$}
& \multicolumn{2}{c|}{$\delta_{uncor}$}
& \multicolumn{2}{c|}{$\delta_{cor}$}
& \multicolumn{2}{c|}{$\delta_{exp}$}
& \multicolumn{2}{c|}{$\delta_{proc}$}
& \multicolumn{2}{c|}{$\delta_{tot}$} \\ 
(GeV$^2$) &  &  &  
& \multicolumn{2}{c|}{(\%)}
& \multicolumn{2}{c|}{(\%)}
& \multicolumn{2}{c|}{(\%)}
& \multicolumn{2}{c|}{(\%)}
& \multicolumn{2}{c|}{(\%)}
& \multicolumn{2}{c|}{(\%)} \\ 
\hline
  15.3 &   0.0562 &   0.0750 &   0.0174 &   7&6 &   1&4 &   4&7 &   9&1 &  2&1 &   9&3 \\
  15.3 &   0.0562 &   0.0900 &   0.0162 &  10&  &   3&6 &   5&1 &  12&  &  2&8 &  12&  \\
  15.3 &   0.1780 &   0.0025 &   0.0136 &   3&4 &   0&5 &   5&0 &   6&0 &  1&3 &   6&2 \\
  15.3 &   0.1780 &   0.0085 &   0.0111 &   3&4 &   0&5 &   4&8 &   5&9 &  2&2 &   6&2 \\
  15.3 &   0.1780 &   0.0160 &   0.0098 &   3&9 &   0&6 &   5&0 &   6&4 &  2&2 &   6&8 \\
  15.3 &   0.1780 &   0.0250 &   0.0097 &   6&1 &   0&9 &   5&2 &   8&1 &  2&4 &   8&4 \\
  15.3 &   0.1780 &   0.0350 &   0.0117 &  15&  &  17&  &   4&9 &  23&  &  2&3 &  23&  \\
  15.3 &   0.1780 &   0.0500 &   0.0134 &  12&  &  15&  &   4&9 &  20&  &  2&3 &  20&  \\
  15.3 &   0.5620 &   0.0009 &   0.0180 &   8&8 &   3&4 &   4&6 &  11&  &  3&3 &  11&  \\
  15.3 &   0.5620 &   0.0025 &   0.0173 &   2&5 &   0&2 &   5&8 &   6&3 &  3&5 &   7&2 \\
  15.3 &   0.5620 &   0.0085 &   0.0162 &   3&3 &   0&5 &   5&1 &   6&1 &  3&0 &   6&8 \\
  15.3 &   0.5620 &   0.0160 &   0.0151 &  17&  &  14&  &   4&9 &  22&  &  2&3 &  22&  \\
  15.3 &   0.5620 &   0.0350 &   0.0094 &  20&  &  21&  &   4&9 &  30&  &  2&3 &  30&  \\
  26.5 &   0.0056 &   0.0750 &   0.0359 &  17&  &   0&0 &   5&3 &  18&  &  3&2 &  18&  \\
  26.5 &   0.0178 &   0.0250 &   0.0179 &   8&0 &   1&4 &   4&8 &   9&4 &  2&3 &   9&7 \\
  26.5 &   0.0178 &   0.0350 &   0.0202 &   8&6 &   0&0 &   5&3 &  10&  &  1&6 &  10&  \\
  26.5 &   0.0178 &   0.0500 &   0.0250 &   6&7 &   1&3 &   4&8 &   8&4 &  1&8 &   8&6 \\
  26.5 &   0.0178 &   0.0750 &   0.0249 &  10&  &   2&3 &   5&2 &  12&  &  2&6 &  12&  \\
  26.5 &   0.0562 &   0.0085 &   0.0157 &   6&6 &   1&2 &   5&3 &   8&6 &  8&0 &  12&  \\
  26.5 &   0.0562 &   0.0160 &   0.0150 &   4&9 &   0&7 &   4&8 &   7&0 &  1&8 &   7&2 \\
  26.5 &   0.0562 &   0.0250 &   0.0134 &   5&5 &   0&7 &   4&5 &   7&1 &  1&3 &   7&3 \\
  26.5 &   0.0562 &   0.0350 &   0.0157 &   7&4 &   0&0 &   4&8 &   8&8 &  1&6 &   9&0 \\
  26.5 &   0.0562 &   0.0500 &   0.0184 &   6&2 &   1&6 &   5&1 &   8&2 &  1&3 &   8&3 \\
  26.5 &   0.0562 &   0.0750 &   0.0211 &   7&4 &   1&8 &   4&5 &   8&9 &  1&5 &   9&0 \\
  26.5 &   0.0562 &   0.0900 &   0.0237 &   9&6 &   3&2 &   5&0 &  11&  &  3&4 &  12&  \\
  26.5 &   0.1780 &   0.0025 &   0.0138 &   5&4 &   0&4 &   5&1 &   7&5 &  1&4 &   7&6 \\
  26.5 &   0.1780 &   0.0085 &   0.0126 &   5&0 &   0&8 &   4&8 &   7&0 &  2&7 &   7&5 \\
  26.5 &   0.1780 &   0.0160 &   0.0113 &   5&5 &   0&0 &   5&1 &   7&6 &  2&2 &   7&9 \\
  26.5 &   0.1780 &   0.0250 &   0.0093 &   6&5 &   1&0 &   4&9 &   8&2 &  1&4 &   8&3 \\
  26.5 &   0.1780 &   0.0350 &   0.0100 &   9&8 &   0&0 &   5&7 &  11&  &  4&0 &  12&  \\
  26.5 &   0.1780 &   0.0500 &   0.0105 &  26&  &  14&  &   4&9 &  30&  &  1&9 &  30&  \\
  26.5 &   0.1780 &   0.0750 &   0.0169 &  42&  &  11&  &   4&9 &  44&  &  1&9 &  44&  \\
  26.5 &   0.5620 &   0.0009 &   0.0241 &  22&  &  10&  &   4&9 &  25&  &  1&9 &  25&  \\
  26.5 &   0.5620 &   0.0025 &   0.0189 &   3&7 &   0&2 &   6&0 &   7&0 &  9&1 &  12&  \\
  26.5 &   0.5620 &   0.0085 &   0.0140 &   4&3 &   0&4 &   5&0 &   6&6 &  3&8 &   7&6 \\
  26.5 &   0.5620 &   0.0250 &   0.0136 &  31&  &  15&  &   4&9 &  35&  &  1&9 &  35&  \\
\hline
\end{tabular}
\captcont{continued}
\label{tab:xsec:d}
\end{center}
\end{table}

\begin{table}[ht!]
\begin{center}
\begin{tabular}{|c|c|c|c|r@{.}l|r@{.}l|r@{.}l|r@{.}l|r@{.}l|r@{.}l|} 
\hline
$Q^2$ & $\beta$ & $x_{I\!\!P}$ & $x_{I\!\!P} \sigma_r^{D(3)}$ 
& \multicolumn{2}{c|}{$\delta_{stat}$}
& \multicolumn{2}{c|}{$\delta_{uncor}$}
& \multicolumn{2}{c|}{$\delta_{cor}$}
& \multicolumn{2}{c|}{$\delta_{exp}$}
& \multicolumn{2}{c|}{$\delta_{proc}$}
& \multicolumn{2}{c|}{$\delta_{tot}$} \\ 
(GeV$^2$) &  &  &  
& \multicolumn{2}{c|}{(\%)}
& \multicolumn{2}{c|}{(\%)}
& \multicolumn{2}{c|}{(\%)}
& \multicolumn{2}{c|}{(\%)}
& \multicolumn{2}{c|}{(\%)}
& \multicolumn{2}{c|}{(\%)} \\ 
\hline
  46   &   0.0178 &   0.0500 &   0.0313 &   8&6 &   4&5 &   4&7 &  11&  &   1&6 &  11&     \\
  46   &   0.0178 &   0.0750 &   0.0218 &  19&  &   0&0 &   5&1 &  20&  &   2&5 &  20&    \\
  46   &   0.0562 &   0.0160 &   0.0163 &   8&8 &   0&0 &   5&2 &  10&  &   2&1 &  11&    \\
  46   &   0.0562 &   0.0250 &   0.0172 &   8&6 &   0&0 &   5&3 &  10&  &   2&1 &  10&    \\
  46   &   0.0562 &   0.0350 &   0.0158 &   8&3 &   1&8 &   4&6 &   9&6 &   2&2 &   9&8   \\
  46   &   0.0562 &   0.0500 &   0.0199 &   7&6 &   1&9 &   4&8 &   9&2 &   2&8 &   9&6   \\
  46   &   0.0562 &   0.0750 &   0.0212 &   8&4 &   1&2 &   4&9 &   9&7 &   3&2 &  10&    \\
  46   &   0.0562 &   0.0900 &   0.0267 &   8&9 &   2&4 &   4&8 &  10&  &   1&0 &  10&    \\
  46   &   0.1780 &   0.0085 &   0.0121 &   6&6 &   1&3 &   5&4 &   8&6 &   2&1 &   8&9   \\
  46   &   0.1780 &   0.0160 &   0.0133 &   5&9 &   1&5 &   4&8 &   7&7 &   2&4 &   8&1   \\
  46   &   0.1780 &   0.0250 &   0.0135 &   8&5 &   0&0 &   4&9 &   9&8 &   2&2 &  10&    \\
  46   &   0.1780 &   0.0350 &   0.0129 &   7&5 &   1&9 &   4&6 &   9&0 &   2&1 &   9&2   \\
  46   &   0.1780 &   0.0500 &   0.0148 &   7&4 &   2&9 &   4&8 &   9&3 &   2&4 &   9&6   \\
  46   &   0.1780 &   0.0750 &   0.0201 &   9&9 &   4&0 &   4&7 &  12&  &   3&4 &  12&    \\
  46   &   0.1780 &   0.0900 &   0.0177 &  13&  &   4&2 &   5&0 &  14&  &   8&6 &  17&    \\
  46   &   0.5620 &   0.0025 &   0.0196 &   5&1 &   1&0 &   5&4 &   7&5 &   4&2 &   8&6   \\
  46   &   0.5620 &   0.0085 &   0.0135 &   5&1 &   1&0 &   4&9 &   7&2 &   4&6 &   8&5   \\
  46   &   0.5620 &   0.0160 &   0.0124 &   6&9 &   1&8 &   4&8 &   8&6 &   2&3 &   8&9   \\
  46   &   0.5620 &   0.0250 &   0.0106 &  13&  &   0&0 &   5&9 &  14&  &   1&2 &  15&    \\
  46   &   0.5620 &   0.0350 &   0.0135 &  14&  &   7&0 &   4&8 &  16&  &   2&2 &  16&    \\
  46   &   0.5620 &   0.0500 &   0.0120 &  17&  &  20&  &   4&9 &  26&  &   2&3 &  26&    \\
  46   &   0.8160 &   0.0009 &   0.0145 &  21&  &   5&3 &   4&5 &  22&  &   1&4 &  22&    \\
  46   &   0.8160 &   0.0025 &   0.0131 &  17&  &   8&1 &   5&3 &  20&  &   3&0 &  20&    \\
  46   &   0.8160 &   0.0085 &   0.0110 &  18&  &   3&9 &   4&3 &  19&  &   1&5 &  19&    \\
  46   &   0.8160 &   0.0160 &   0.0092 &  27&  &   3&9 &   5&4 &  28&  &   4&1 &  28&    \\
  80   &   0.0562 &   0.0350 &   0.0227 &  19&  &   0&0 &   5&8 &  20&  &   2&7 &  20&    \\
  80   &   0.0562 &   0.0500 &   0.0235 &  15&  &   0&0 &   5&0 &  16&  &   2&0 &  16&    \\
  80   &   0.0562 &   0.0750 &   0.0216 &  24&  &   0&0 &   5&9 &  25&  &   1&9 &  25&    \\
  80   &   0.1780 &   0.0085 &   0.0206 &  15&  &   0&0 &   6&0 &  16&  &   2&9 &  16&    \\
  80   &   0.1780 &   0.0160 &   0.0133 &  13&  &   0&0 &   4&8 &  14&  &   2&3 &  14&    \\
  80   &   0.1780 &   0.0250 &   0.0146 &  12&  &   0&0 &   5&2 &  13&  &   1&6 &  13&    \\
  80   &   0.1780 &   0.0350 &   0.0162 &  14&  &   0&0 &   5&6 &  15&  &   1&0 &  15&    \\
  80   &   0.1780 &   0.0500 &   0.0146 &  15&  &   0&0 &   5&5 &  16&  &   2&3 &  16&    \\
  80   &   0.1780 &   0.0750 &   0.0183 &  26&  &   0&0 &   5&3 &  27&  &   3&0 &  27&    \\
  80   &   0.5620 &   0.0085 &   0.0116 &  10&  &   0&0 &   6&4 &  12&  &   5&1 &  13&    \\
  80   &   0.5620 &   0.0160 &   0.0090 &  14&  &   0&0 &   7&0 &  15&  &   3&5 &  16&    \\
\hline
\end{tabular}
\captcont{continued}
\label{tab:xsec:e}
\end{center}
\end{table}

\begin{table}[t!]
\begin{center}
\begin{tabular}{|c|c|c|c|r@{.}l|r@{.}l|r@{.}l|r@{.}l|r@{.}l|r@{.}l|} 
\hline
$Q^2$ & $\beta$ & $x_{I\!\!P}$ & $x_{I\!\!P} \sigma_r^{D(3)}$ 
& \multicolumn{2}{c|}{$\delta_{stat}$}
& \multicolumn{2}{c|}{$\delta_{uncor}$}
& \multicolumn{2}{c|}{$\delta_{cor}$}
& \multicolumn{2}{c|}{$\delta_{exp}$}
& \multicolumn{2}{c|}{$\delta_{proc}$}
& \multicolumn{2}{c|}{$\delta_{tot}$} \\ 
(GeV$^2$) &  &  &  
& \multicolumn{2}{c|}{(\%)}
& \multicolumn{2}{c|}{(\%)}
& \multicolumn{2}{c|}{(\%)}
& \multicolumn{2}{c|}{(\%)}
& \multicolumn{2}{c|}{(\%)}
& \multicolumn{2}{c|}{(\%)} \\ 
\hline
  80   &   0.5620 &   0.0250 &   0.0104 &  17&  &   0&0 &   6&7 &  18&  &   5&3 &  19&  \\
  80   &   0.5620 &   0.0350 &   0.0109 &  25&  &   0&0 &   7&3 &  26&  &   3&6 &  26&  \\
 200   &   0.0562 &   0.0500 &   0.0162 &  28&  &   0&0 &   5&0 &  28&  &   1&0 &  28&  \\
 200   &   0.0562 &   0.0750 &   0.0288 &  37&  &   0&0 &   5&5 &  37&  &   2&3 &  37&  \\
 200   &   0.1780 &   0.0160 &   0.0145 &  20&  &   0&0 &   5&8 &  21&  &   1&3 &  21&  \\
 200   &   0.1780 &   0.0250 &   0.0199 &  16&  &   0&0 &   5&0 &  17&  &   1&9 &  17&  \\
 200   &   0.1780 &   0.0350 &   0.0169 &  22&  &   0&0 &   5&2 &  23&  &   2&6 &  23&  \\
 200   &   0.1780 &   0.0500 &   0.0235 &  20&  &   0&0 &   5&5 &  21&  &   2&6 &  21&  \\
 200   &   0.1780 &   0.0750 &   0.0209 &  35&  &   0&0 &   5&6 &  35&  &   2&5 &  36&  \\
 200   &   0.5620 &   0.0085 &   0.0109 &  19&  &   0&0 &   6&6 &  21&  &   3&9 &  21&  \\
 200   &   0.5620 &   0.0160 &   0.0093 &  23&  &   0&0 &   6&4 &  24&  &   1&9 &  24&  \\
 200   &   0.5620 &   0.0250 &   0.0074 &  27&  &   0&0 &   6&7 &  28&  &   4&9 &  29&  \\
 200   &   0.5620 &   0.0350 &   0.0158 &  33&  &   0&0 &   6&7 &  34&  &   2&4 &  34&  \\
 200   &   0.5620 &   0.0500 &   0.0151 &  29&  &   0&0 &   5&4 &  29&  &   1&8 &  29&  \\
 200   &   0.5620 &   0.0750 &   0.0228 &  50&  &   0&0 &   5&9 &  50&  &   3&2 &  50& \\
\hline
\end{tabular}
\captcont{continued}
\label{tab:xsec:f}
\end{center}
\end{table}


\end{document}